\documentclass[twocolumn]{aastex701}
\graphicspath{{./}{}{figures/}}
\usepackage[T1]{fontenc}
\usepackage{multirow}
\usepackage{booktabs}
\usepackage{amsmath}

\begin{document}

\title{exoALMA XXV: A Statistical Analysis of Gaseous Substructure in Low-Inclination Disks}

\author[0000-0003-1534-5186]{Richard Teague} 
\affiliation{Department of Earth, Atmospheric and Planetary Sciences, Massachusetts Institute of Technology, Cambridge, MA 02139, USA}
\email{rteague@mit.edu}

\author[0009-0000-7872-3493]{Sean M. Andrews}
\affiliation{Center for Astrophysics | Harvard \& Smithsonian, Cambridge, MA 02138, USA}
\email[]{}

\author[0000-0001-7258-770X]{Jaehan Bae}
\affiliation{Department of Astronomy, University of Florida, Gainesville, FL 32611, USA}
\email[]{}

\author[0000-0001-6378-7873]{Marcelo Barraza-Alfaro}
\affiliation{Department of Earth, Atmospheric and Planetary Sciences, Massachusetts Institute of Technology, Cambridge, MA 02139, USA}
\affiliation{Niels Bohr International Academy, Niels Bohr Institute, Blegdamsvej 17, DK-2100 Copenhagen Ø, Denmark}
\email[]{}

\author[0000-0002-7695-7605]{Myriam Benisty}
\affiliation{Max-Planck Institute for Astronomy (MPIA), Königstuhl 17, 69117 Heidelberg, Germany}
\email[]{}

\author[0000-0002-1483-8811]{Ian Czekala} 
\affiliation{Australia Telescope National Facility, CSIRO Space and Astronomy, PO Box 76, Epping, NSW 1710, Australia}
\email{ian.czekala@csiro.au}

\author[0000-0003-4689-2684]{Stefano Facchini}
\affiliation{Dipartimento di Fisica, Universit\'a degli Studi di Milano, Via Celoria 16, 20133 Milano, Italy}
\email[]{}

\author[0000-0003-4679-4072]{Daniele Fasano}
\affiliation{INAF, Osservatorio Astronomico di Roma, Via Frascati 33, 00078, Monte Porzio Catone, Italy}
\email[]{}

\author[0000-0002-9298-3029]{Mario Flock}
\affiliation{Max-Planck Institute for Astronomy (MPIA), Königstuhl 17, 69117 Heidelberg, Germany}
\email[]{}

\author[0000-0003-1117-9213]{Misato Fukagawa} 
\affiliation{Astronomical Institute, Graduate School of Science, Tohoku University, 6-3 Aoba, Aramaki, Aoba-ku, Sendai, Miyagi 980-8578 Japan}
\email[]{}

\author[0000-0002-8138-0425]{Cassandra Hall}
\affiliation{Center for Simulational Physics, The University of Georgia, Athens, GA 30602, USA}
\email[]{}

\author[0000-0001-8446-3026]{Andr\'es F. Izquierdo} 
\affiliation{Department of Astronomy, University of Florida, Gainesville, FL 32611, USA}
\affiliation{NASA Hubble Fellowship Program Sagan Fellow}
\email[]{}

\author[0000-0002-2357-7692]{Giuseppe Lodato} 
\affiliation{Dipartimento di Fisica, Universit\'a degli Studi di Milano, Via Celoria 16, 20133 Milano, Italy}
\email[]{}

\author[0000-0002-0491-143X]{Jochen Stadler} 
\affiliation{European Southern Observatory, Karl-Schwarzschild-Str. 2, 85748 Garching bei München, Germany}
\email[]{}

\author[0000-0002-8590-7271]{Jason P. Terry}
\affiliation{Department of Earth Sciences, University of Oxford}
\email{jason.terry@earth.ox.ac.uk}  

\author[0000-0003-1526-7587]{David J. Wilner}
\affiliation{Center for Astrophysics | Harvard \& Smithsonian, Cambridge, MA 02138, USA}
\email[]{}

\author[0000-0002-7501-9801]{Andrew J. Winter}
\affiliation{Astronomy Unit, School of Physics and Astronomy, Queen Mary University of London, London E1 4NS, UK}
\email[]{}

\author[0000-0001-8002-8473]{Tomohiro C. Yoshida}
\affiliation{Dipartimento di Fisica, Universit\'a degli Studi di Milano, Via Celoria 16, 20133 Milano, Italy}
\email[]{}

\begin{abstract}
Protoplanetary disks exhibit complex structure in both their gas dynamics and their temperature. Characterizing it has proven challenging, owing to the mismatch between the observed complexity and the relatively simple semi-analytical models typically used to interpret it. To augment these analyses, we apply the second-order structure function to the velocity and temperature perturbations of four low-inclination sources from the exoALMA program, each measured against a noise-only null built from the data themselves. All exhibit correlated structure. Its radial correlation length is limited by the beam in every case, which excludes non-axisymmetric structure on radial scales larger than ${\sim}1.5$ pressure scale heights. In three of the four disks the structure is instead resolved and elongated in azimuth, by factors of $1.6$--$2.3$, as expected of small-scale perturbations sheared out by Keplerian rotation. We further introduce two statistics that characterize spiral structure without tracing an arm crest: a \emph{modal power} spectrum, measuring the periodicity of the azimuthal structure, and a \emph{winding contrast}, measuring the radial dependence of its phase. The velocity residuals of three disks are significantly wound. MWC~758 is the only source in which the winding is coherent across the disk and shared by the temperature field. Its trailing $m=1$ temperature structure is consistent with the illumination of a warp, while its velocity field carries small-scale, sheared structure that could redistribute angular momentum. Because neither statistic requires a crest, both extend a quantitative description of winding to the flocculent regime, where a spiral cannot be parameterized.
\end{abstract}

\section{Introduction} 

High-resolution ALMA observations have established that the gas in a protoplanetary disk does not simply rotate in a Keplerian fashion. Once an axisymmetric model is subtracted, the line-of-sight velocity retains residuals at the few per cent level relative to the background Keplerian rotation, and the emission surface retains temperature variations of a few Kelvin, ${\approx}5$--$10\%$ of the azimuthally averaged surface temperature, both organized on scales comparable to the disk's own radial extent \citep{Teague_ea_2021}. The exoALMA program, \citet{Teague_ea_2025}, was designed to characterize exactly this component, and its atlases of non-Keplerian deviations and of vertical flows found spiral-like features, arcs, rings and localized flows across the majority of the sample \citep{Fukagawa_ea_2026, Benisty_ea_2026}.

Interpreting that residual component is harder than detecting it. The residual velocity field is usually compared against a parametric expectation (the wake of an embedded planet, a warp, a vortex, a gravitationally unstable arm) and individual disks have been interpreted convincingly in each of these ways \citep[e.g.,][]{Pinte_ea_2018, Pinte_ea_2019, Casassus_Perez_2019, Rosenfeld_ea_2014, Winter_ea_2025, Wolfer_ea_2025, Hall_ea_2020, Longarini_ea_2021, Speedie_ea_2024, Stadler_ea_2023, Izquierdo_ea_2026}. Two difficulties arise. The first is that these models are not easily distinguished from one another on the basis of a residual map, since each predicts a pattern of comparable amplitude on comparable scales. The second, and the one that motivates this work, is that a substantial part of what the data show cannot be described by these parametric expectations. Although the structure is correlated over a finite range of scales, it is neither periodic, ordered, nor traceable as a crest.

A complementary approach is to describe the field statistically rather than parametrically. The second-order structure function, $S_2$, measures the mean square difference of a field between pairs of points as a function of their separation, and so characterizes the amplitude and the characteristic scales of a correlated field without reference to any particular morphology. It has a long history in the study of turbulence \citep{Kolmogorov_1941, Monin_Yaglom_1975}, and has been applied extensively to the interstellar medium \citep{Larson_1981, Ossenkopf_MacLow_2002, Heyer_Brunt_2004, Elmegreen_Scalo_2004} and the solar wind \citep{Bruno_Carbone_2013}, including specifically to velocity centroid maps of the kind we analyze here \citep{Esquivel_Lazarian_2005}. Its appeal in the present context is that it makes no assumption of a crest, a mode number or a pitch angle, and returns a measurement whether or not the field is ordered.

\citet{Teague_ea_2026} developed this framework for disk velocity residuals and established what can and cannot be recovered from noisy data. Structure is detectable well below the amplitude at which it can be determined by eye, and the dominant correlation scales and the degree of anisotropy can be constrained, provided that significance is assessed against a noise model with properties matched to the data rather than against an idealized noise model. Application to numerical simulations of the vertical shear instability, the gravitational instability and the magneto-rotational instability demonstrated that this analysis was able to differentiate between models, however not uniquely identify them. This application also identified the framework's principal limitation: a wound field, i.e., a spiral, imprints its winding on the off-diagonal structure of $S_2$, where an annulus-by-annulus analysis, as used to characterize the scale of the perturbations, averages it away. In effect, the presence of spiral structure dilutes the significance of the scale parameters.

This paper applies that methodology to four near-face-on exoALMA sources, J1604, HD~135344B, HD~143006 and MWC~758, extending the analysis to temperature fluctuations in addition to velocity perturbations, and introduces new tools for better characterizing the winding in the data. The structure of the paper is as follows. Section~\ref{sec:substructures} presents the velocity and temperature perturbation fields for the four sources, including the development of appropriate noise models. The $S_2$ analysis is performed in Section~\ref{sec:structure_function} while Section~\ref{sec:spirals} presents and applies new techniques to characterize the winding of the fields. The results are interpreted in Section~\ref{sec:discussion} in terms of the radial scale, the shear and the winding of the structure, with the main findings summarized in Section~\ref{sec:conclusions}.

\section{Velocity and Intensity Substructures}
\label{sec:substructures}

A spectrally resolved line records information regarding a variety of properties of the emitting gas, including the bulk line-of-sight velocity, the velocity dispersion and the gas temperature \citep[e.g.,][]{Izquierdo_ea_2025}. \citet{Izquierdo_ea_2026} demonstrates how the combined analyses of these components can offer insights into the dynamical processes that are driving the substructures that we see. For this paper we focus on two commonly used properties: the line center, mapping the bulk velocity of the gas, and the line peak, a measure of the gas temperature when the emission is optically thick \citep[almost guaranteed for $^{12}$CO emission;][]{Galloway-Sprietsma_ea_2025}.

\subsection{Sources}
\label{sec:substructures:sources}

Owing to the complexities associated with analyzing the emission from moderately inclined disks where the front- and back-sides are distinguishable, we focus this work on the four least inclined exoALMA sources: J1604, HD~135344B, HD~143006 and MWC~758. All satisfy $|i| < 20\degr$. For the following analysis, we adopt the inclination and position angles described in exoALMA I \citep{Teague_ea_2025} and re-derive source centers and dynamical masses as reproduced in Table~\ref{tab:sources} which have been found to be sensitive to the method used to produce the velocity map \citep{Teague_Foreman-Mackey_2018}. To calculate the moment maps in the following sections, we selected the `fiducial' images which were imaged with a $100~{\rm m\,s^{-1}}$ channel spacing and restored with a circular CLEAN beam with a FWHM of $0\farcs15$ \citep{Loomis_ea_2025}.

\begin{figure*}
    \centering
    \includegraphics[]{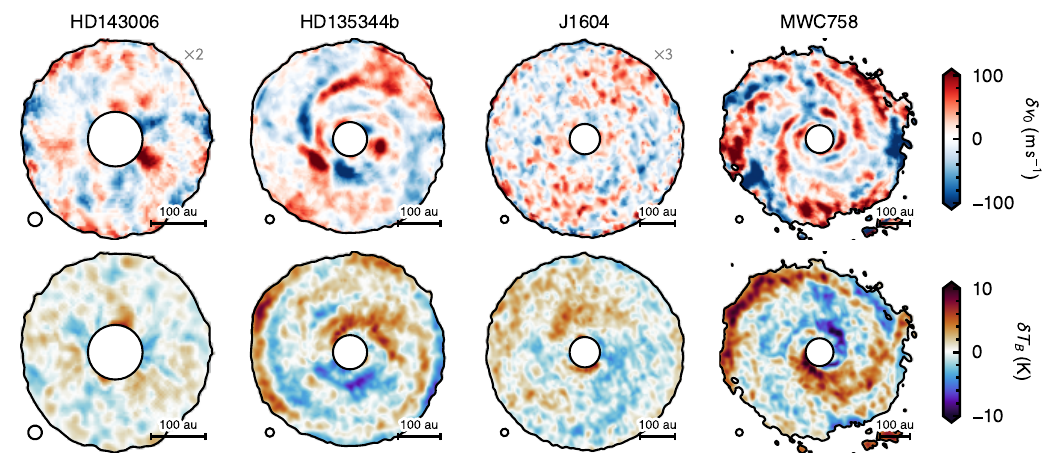}
    \caption{The on-sky projection of the velocity residuals (top) and peak brightness temperature residuals (bottom) for the four sources. The black circle in the lower left shows the $0\farcs15$ CLEAN beam and the scale bar in the lower right shows 100~au. Note that for HD~143006 and J1604, the velocity residuals have been inflated by a factor of two and three, respectively, such that they are visible, while additionally HD~143006 has had its shadow removed in the peak brightness temperature residual. All maps are masked where the peak brightness temperature drops below 7~K. All four disks rotate clockwise on the sky bar HD~135344B, which rotates counter-clockwise \citep[][see Table~\ref{tab:sources}]{Izquierdo_ea_2025}.}
    \label{fig:onsky_residuals}
\end{figure*}

\begin{figure*}
    \centering
    \includegraphics[width=\textwidth]{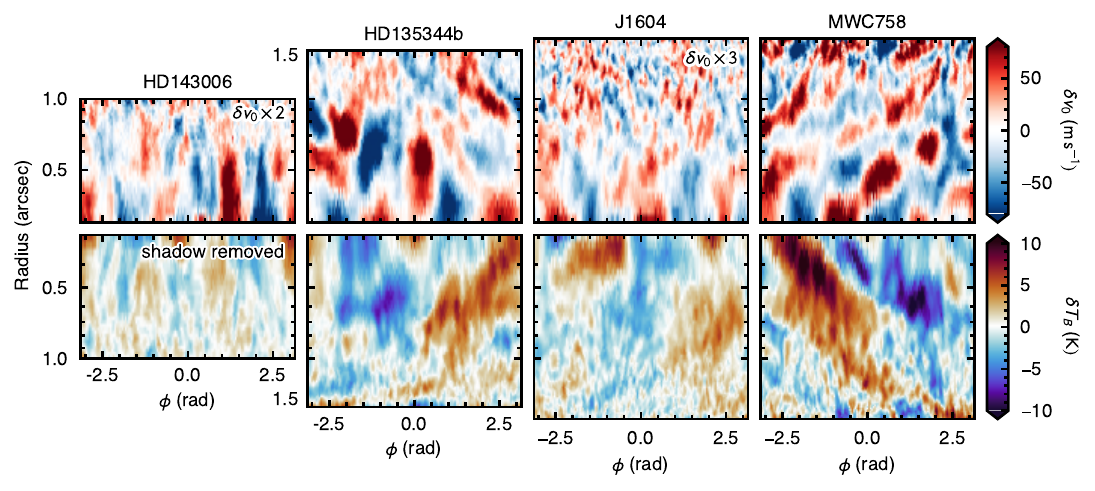}
    \caption{Velocity (top) and temperature (bottom) residuals for the four disks shown in a polar projection. Note that the bottom row is flipped upside down. The inner radius for all panels is $0\farcs3$, twice the beam major axis and the radial axis is shown in log-space to highlight logarithmic spiral structure. The velocity residuals for HD~143006 and J1604 have been inflated by a factor of two and three, respectively, to make them visible with this color scale.}
    \label{fig:polar_residuals}
\end{figure*}

\subsection{Velocity Maps}
\label{sec:substructures:velocity}

To calculate the velocity maps we use \texttt{bettermoments} \citep{Teague_Foreman-Mackey_2018} to collapse the data using the \texttt{percentiles} method (essentially an intensity-weighted \emph{median} velocity, rather than the intensity-weighted \emph{average} velocity of the traditional first moment) after applying a third-order Savitzky-Golay filter across a five-sample window. We found that this method recovers robust velocity maps even when faced with noisy data or asymmetric line profiles.

We subtract a background model comprising azimuthally symmetric annuli that contain $v_\phi$, $v_r$ and $v_z$ components using the \texttt{fit\_annuli} function in \texttt{eddy}. This approach is similar to the method used in \citet{Casassus_Perez_2019} but holding the disk inclination and position angles fixed across all radii, and differs from the \texttt{discminer} approach described in \citet{Izquierdo_ea_2025} and used in \citet{Fukagawa_ea_2026}. We opted for this approach as our goal is to remove as much large-scale structure from the velocity maps as possible. This radial decomposition has the flexibility to mimic the single-armed spiral morphology that \citet{Winter_ea_2025} interpreted as a signature of a moderate warp in the disk.

An important aspect of this model subtraction that has importance later is that despite the velocity model being intrinsically azimuthally symmetric ($m=0$), the projection of the velocity along the line of sight transforms this to an $m=1$ mode and so most of the $m=1$ power is removed from the residuals. Only the axisymmetric part of each component is removed, however, so the residual retains the non-axisymmetric components. The resulting velocity residuals are shown in Figure~\ref{fig:onsky_residuals} and in a log-polar projection in Figure~\ref{fig:polar_residuals}. Note that the velocity residuals for HD~143006 and J1604 have been increased by a factor of two and three, respectively, to make them visible on the joint color scale.

\subsection{Peak Brightness Temperature Maps}
\label{sec:substructures:intensity}

For the peak brightness temperature, we used the standard \texttt{quadratic} method in \texttt{bettermoments}, applying the same smoothing as for the velocity maps and converting to brightness temperature using the Rayleigh-Jeans approximation. We adopt this rather than the full Planck law because the conversion is then linear in $I_\nu$, so the residuals we characterize are a constant rescaling of those measured in ${\rm Jy\,beam^{-1}}$, and because it remains well behaved at the low intensities of the outer disk. The only impact on the analysis that this has is that the amplitude of the variations derived may underestimate the true temperature deviations. We use \texttt{GoFish} \citep{GoFish} to calculate an azimuthally averaged radial profile of the peak intensity and use this to create a smooth background model which is subtracted from the peak intensity map. The resulting variations are shown in the bottom row of Figure~\ref{fig:onsky_residuals} and Figure~\ref{fig:polar_residuals}.

As HD~143006 has a prominent shadow which dominates the peak brightness temperature \citep{Benisty_ea_2018}, we removed a simple shadow model from each radius. The model is a sinusoidal function, where the amplitude and phase are allowed to vary, but the period is fixed to $2\pi$, that was optimized for using \texttt{scipy.optimize.curve\_fit}. As can be seen in Figure~\ref{fig:polar_residuals}, the subtraction appears to remove most structure from the residual. Unless stated otherwise, every HD~143006 peak brightness temperature result below is measured on this shadow-subtracted residual, and its matched null is rebuilt with the identical subtraction, so that observation and null are always filtered alike.

\subsection{Noise Models}
\label{sec:substructures:noise_models}

To build the noise models we calculated a noise-only structure function $S_2^{\rm noise}$ based on the first and last ten (all signal-free) channel maps of each dataset. As discussed in Appendix~\ref{sec:app:noise_models}, we found this to be a more faithful representation of the noise than an analytical $S_2$ constructed using the CLEAN beam properties and the variance of the noise-only channels as discussed in \citet{Teague_ea_2026}. We drew many realizations of noise which were used to corrupt a simple, morphological model of the disk emission based on the observational data which was subsequently passed through the analysis pipeline. The resulting velocity and peak brightness temperature residuals comprised the null sample for each source and property, with the HD~143006 shadow model of Section~\ref{sec:substructures:intensity} refitted and removed from every draw exactly as it is from the observation. Appendix~\ref{sec:app:noise_models} provides more details on this process.

\section{Structure Function Analysis}
\label{sec:structure_function}

With residual maps for both properties and a matched noise ensemble for each, we can now apply the framework of \citet{Teague_ea_2026} identically to all eight (source, property) combinations. The second order structure function of field $f$ is defined as
\begin{equation}
    S_2(\ell) = \langle |f(\mathbf{x}) - f(\mathbf{x} + \ell)|^2 \rangle,
\end{equation}
where the lag, $\ell$, describes the scalar distance between two points within that field and the angle brackets denote an average over all $\ell$. For application to protoplanetary disks we consider radial and azimuthal lags separately: $S_2(\delta r, \delta\phi)$.

\subsection{Deprojected $S_2$ Stacks}
\label{sec:structure_function:stacks}

Structure functions are calculated with the \texttt{calculate\_structure\_function\_stack} method of \texttt{eddy}. Following the nomenclature from \citet{Teague_ea_2026} we use \emph{global} $S_2$ to describe a structure function calculated without defining a specific reference radius while an $S_2$ \emph{stack} is an ensemble of $S_2$ functions that step through each radius in the disk and use that as the reference radius. The residual map is first deprojected onto a polar $(r,\phi)$ grid using the geometry of Table~\ref{tab:sources}. We adopt a $300 \times 300$ grid spanning $r_{\rm in} \le r \le r_{\rm out}$ and the full $2\pi$ in azimuth with linear interpolation, and step the reference radius across the same range in increments of one pixel with a band half-width of one pixel. The result is a set of two-dimensional $S_2(\delta r, \delta\phi)$ surfaces, one per annulus, with radial lag in arcseconds and azimuthal lag in degrees. We do not symmetrize the radial lag axis, so the inward and outward halves remain separable. Azimuthal lags are retained over the full $\pm 180\degr$, but radial lags are capped at $\delta r \le 0\farcs5$. A cap of $0\farcs5$ is several beams in size and roughly five times the correlation lengths recovered below, so the plateau is always estimated from lags that the data sample fully. This results in between 100 and 400 independent beams across each source: sufficient for a robust determination of $S_2$. 

The identical process, with the same grids and the same geometry, is applied to the observation and to every null draw of Appendix~\ref{sec:app:noise_models}. The extraction biases of Section~\ref{sec:substructures} are therefore common to both and largely cancel in the comparison rather than needing to be modeled explicitly. The polar-coordinate representation additionally imprints an effective azimuthal non-stationarity (i.e., the azimuthal properties change across the field) on $S_2$ itself \citep{Teague_ea_2026}, but this does not propagate into the heuristics: because $\hat{\alpha}_\phi$ is measured on the native angular half-power width rather than on the arc length, \citet{Teague_ea_2026} recover it without bias, and we adopt the same convention here. The background model subtraction described in Section~\ref{sec:substructures} will never introduce structure into $S_2$, but may over-subtract power, potentially suppressing the signal.

\subsection{Heuristics and Significance}
\label{sec:structure_function:heuristics}

\begin{figure}
    \includegraphics[width=0.95\columnwidth]{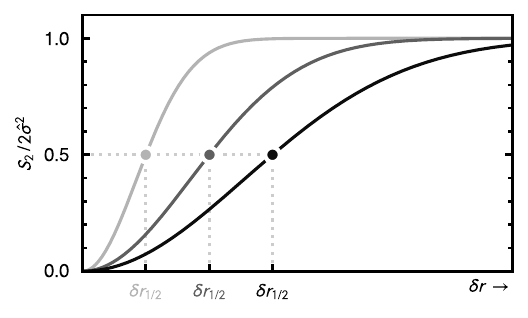}
    \caption{Demonstrating the relationship between the empirical heuristics and $S_2$ with three radial structure functions measured with different reference radii and returning differing correlation lengths. All three profiles have been normalized by their plateau, $2\hat{\sigma}^2$, such that $S_2 / 2\hat{\sigma}^2 = 1$ at large lag. The half-power lag, $\delta r_{1/2}$, is the lag at which $S_2$ reaches half the plateau, or 1/2 with this normalization, and the correlation length is $\hat{\ell}_r = \delta r_{1/2} / \sqrt{2 \ln 2}$. $\hat{\alpha}_r$ quantifies the change in $\hat{\ell}_r$ as the reference radius changes.}
    \label{fig:s2_schematic}
\end{figure}

From each stack we measure the six heuristics of \citet{Teague_ea_2026}, $\{\hat{\sigma},\, \hat{\ell}_r,\, \hat{\ell}_s,\, \hat{\mathcal{A}},\, \hat{\alpha}_r,\, \hat{\alpha}_\phi\}$, with the \texttt{calculate\_heuristics} method. As in \citet{Teague_ea_2026}, they describe three properties of the field. The amplitude, $\hat{\sigma}$, measures how much correlated structure there is: $S_2$ rises to a plateau of $2\hat{\sigma}^2$ at lags beyond the correlation length. The correlation lengths, $\hat{\ell}_r$ and $\hat{\ell}_s$, measure the size of the correlated fluctuations in radius and in azimuth, the latter as a physical arc length, and their ratio,
\begin{equation}
    \hat{\mathcal{A}} = \hat{\ell}_s / \hat{\ell}_r,
\end{equation}
measures their shape: $\hat{\mathcal{A}} = 1$ for isotropic structure and $\hat{\mathcal{A}} > 1$ for structure elongated in azimuth. The stationarity slopes describe how these scales change across the disk, assuming power laws
\begin{equation}
    \ell_r(r) \propto r^{\hat{\alpha}_r}, \qquad \ell_\phi(r) \propto r^{\hat{\alpha}_\phi},
\end{equation}
where $\ell_\phi$ is the angular azimuthal scale, so that $\hat{\alpha}_r = 0$ is a radially constant scale and $\hat{\alpha}_\phi = -1$ a fixed arc length. Each is measured annulus by annulus from the half-power lags of $S_2$ and averaged over annuli, weighted by the number of independent patches each samples, with $\hat{\mathcal{A}}$ taken as the average of the per-annulus ratios.

Following \citet{Teague_ea_2026}, both correlation lengths are quoted as the Gaussian kernel scale $\ell$, the standard deviation of the covariance rather than a lag. The heuristics measure the half-power lag $\delta_{1/2}$, the separation at which $S_2$ first reaches half its plateau, which assumes nothing about the shape of the kernel; we divide it by $\sqrt{2 \ln 2}$ to report an $\ell$, so that a Gaussian beam of FWHM $\theta_b$ has $\ell = \theta_b / 2\sqrt{\ln 2} = 0.60\,\theta_b$ and the $0\farcs15$ beam of these images has $\ell = 0\farcs090$. Every length quoted below, in Table~\ref{tab:heuristics} and in the appendices alike, is an $\ell$ in this convention; the anisotropy and the two stationarity slopes are a ratio and two log-log slopes of the underlying lags, and so are independent of it.

\begin{figure}
    \includegraphics[width=\columnwidth]{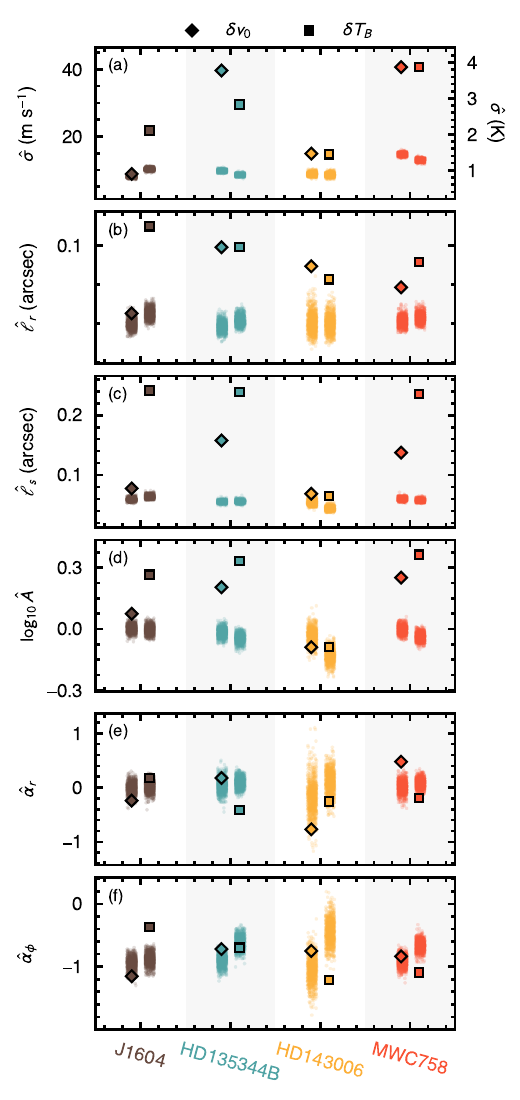}
    \caption{The extracted heuristics for each of the four sources based on the velocity residuals (diamonds) and the peak brightness temperature (squares). Each cloud of points is the source- and property-matched noise-only null of Appendix~\ref{sec:app:noise_models}, $N=1{,}000$ draws carried through the identical extraction; its center is the value that noise alone produces and its width sets the scale against which the observed point is significant. Note that the top row carries two vertical axes: velocity on the left in ${\rm m\,s^{-1}}$ and peak brightness temperature on the right in K.}
    \label{fig:heuristic_extraction}
\end{figure}

Significance is calculated as a $z$-score against the source- and property-matched noise-only ensemble,
\begin{equation}
    z_\theta = \frac{\hat{\theta}_{\rm obs} - {\rm med}(\hat{\theta}_{\rm null})}{\Delta \hat{\theta}_{\rm null}},
\end{equation}
where $\Delta \hat{\theta}_{\rm null}$ is the $16$th--$84$th percentile half-width of the null taken on the side of the median on which the observation falls, so that a skewed null is not mis-calibrated. The anisotropy is compared in $\log_{10} \hat{\mathcal{A}}$ rather than linearly, as its null distribution is close to log-normal and a linear comparison is dominated by the positive tail. Each null comprises the $N = 1{,}000$ draws of Appendix~\ref{sec:app:noise_models}, with each draw plotted in Figure~\ref{fig:heuristic_extraction}.

\subsection{Results}
\label{sec:structure_function:results}

\centerwidetable
\begin{deluxetable*}{l*{6}{@{\hspace{7pt}}c}}
\renewcommand{\arraystretch}{1.1}
\tabletypesize{\footnotesize}
\tablecaption{Structure function heuristics and their significance against the matched noise-only null.\label{tab:heuristics}}
\tablehead{
    \colhead{Source} & \colhead{$\hat{\sigma}$} & \colhead{$\hat{\ell}_r$ ($\arcsec$)} &
    \colhead{$\hat{\ell}_s$ ($\arcsec$)} & \colhead{$\hat{\mathcal{A}}$} &
    \colhead{$\hat{\alpha}_r$} & \colhead{$\hat{\alpha}_\phi$}
}
\startdata
\multicolumn{7}{c}{Velocity Residuals} \\[1pt]\hline
J1604      & \phn9 (\phn$+3.2$) & 0.07 (\phn$+2.0$) & 0.08 (\phn\phn$+7.9$) & 1.19 (\phn$+4.3$) & $-0.24$ (\phn$-2.2$) & $-1.15$ ($-2.4$) \\
HD~135344B & 40 ($+93.5$) & 0.10 ($+15.5$) & 0.16 (\phn$+48.4$) & 1.60 ($+11.8$) & $+0.18$ (\phn$+1.1$) & $-0.72$ ($+2.0$) \\
HD~143006  & 15 ($+14.0$) & 0.09 (\phn$+5.8$) & 0.07 (\phn\phn$+3.2$) & 0.81 (\phn$-1.6$) & $-0.77$ (\phn$-2.2$) & $-0.75$ ($+1.6$) \\
MWC~758    & 41 ($+66.4$) & 0.08 (\phn$+7.7$) & 0.14 (\phn$+34.6$) & 1.78 ($+14.8$) & $+0.48$ (\phn$+4.6$) & $-0.84$ ($+1.0$) \\
\hline
\multicolumn{7}{c}{Temperature Residuals} \\[1pt]\hline
J1604      & \phn2 ($+33.6$) & 0.11 ($+18.4$) & 0.24 (\phn$+66.7$) & 1.84 ($+14.3$) & $+0.17$ (\phn$+1.9$) & $-0.37$ ($+5.8$) \\
HD~135344B & \phn3 ($+70.7$) & 0.10 ($+15.8$) & 0.24 (\phn$+82.3$) & 2.14 ($+19.9$) & $-0.42$ (\phn$-5.5$) & $-0.70$ ($-0.8$) \\
HD~143006  & \phn1 ($+13.0$) & 0.08 (\phn$+5.4$) & 0.06 (\phn\phn$+6.2$) & 0.81 (\phn$+1.3$) & $-0.26$ (\phn$-1.9$) & $-1.21$ ($-4.0$) \\
MWC~758    & \phn4 ($+66.1$) & 0.09 ($+12.5$) & 0.24 (\phn$+90.1$) & 2.31 ($+22.6$) & $-0.19$ (\phn$-3.3$) & $-1.09$ ($-5.2$) \\
\enddata
\tablecomments{Each cell gives the observed heuristic with its signed $z$-score against the matched empirical noise-only null in parentheses. Heuristics are measured over $r_{\rm in} \le r \le r_{\rm out}$ as listed in Table~\ref{tab:sources}, with radial lags capped at $0\farcs5$. In the final column, $\hat{\alpha}_\phi = -1$ signifies a radially fixed arclength.}
\end{deluxetable*}

Table~\ref{tab:heuristics} collects the heuristics and their significances, and Figure~\ref{fig:heuristic_extraction} shows them against the null distributions from which those significances are drawn.

Structure, characterized by $\hat\sigma$, is detected in every one of the eight combinations. The velocity residuals of J1604 show the weakest structure constraints, with $z_{\hat{\sigma}} = 3$, while the velocity perturbations of HD~135344B exhibit the strongest, $z_{\hat{\sigma}} = 94$.

We find that the radial correlation length is limited by the beam in every case. While the matched nulls return $\hat{\ell}_r \approx 0\farcs06$, the measured $\hat{\ell}_r$ span $0\farcs07$--$0\farcs11$. Every $\hat{\ell}_r$ exceeds its own null, which establishes that correlated signal is present, but none resolves the radial structure of that signal by more than ${\sim}20\%$. The radial stationarity slopes agree: a correlation length set by the beam is the same at every radius, and $|\hat{\alpha}_r| < 1$ throughout. We therefore treat $\hat{\ell}_r$ as an upper limit on the radial scale of the structure, and return to what it excludes in Section~\ref{sec:discussion:radial}.

The arc length, by contrast, is resolved. Consistent with \citet{Teague_ea_2026}, $\hat{\ell}_s$ is the most significantly detected heuristic in the sample, and in HD~135344B, MWC~758 and the temperature residual of J1604 it reaches $0\farcs14$--$0\farcs24$, well above the beam. This leads to a robust detection of anisotropy: $\hat{\mathcal{A}} > 1$ in six of the eight combinations, and $\hat{\mathcal{A}} = 1.6$--$2.3$ in five of them. Because $\hat{\ell}_r$ is an upper limit, these anisotropies are lower limits. The velocity residual of J1604 is only mildly elongated, and both fields of HD~143006 have $\hat{\mathcal{A}} < 1$ but are consistent with their own nulls, and so are isotropic within the noise.

The azimuthal stationarity slope is the least informative of the six heuristics, as \citet{Teague_ea_2026} anticipated. Most residuals have $\hat{\alpha}_\phi$ near $-1$, i.e., azimuthal structure of roughly fixed physical arc length. The departures from the null at $|z| > 3$ are confined to the peak brightness temperature, whose nulls are not themselves stationary. Noise passed through the peak-intensity extraction acquires apparent azimuthal non-stationarity, and in HD~143006 an apparent radial elongation, whereas the velocity nulls remain isotropic and stationary.

\section{Spiral Characterization}
\label{sec:spirals}

It is clear from Figure~\ref{fig:onsky_residuals} that the velocity and intensity residuals exhibit spiral structure. This is already hinted at with the robust detection of azimuthal structure via the $\hat{\ell}_s$ heuristic described in Section~\ref{sec:structure_function}. As this does not provide a robust characterization of the spiral morphology, we extend our analysis of $S_2$ to characterize the spiral structure in the data. We explore two analyses, namely the modal power and the winding contrast, testing the periodicity of the structure and how `wound' the structure is, respectively. These use the \emph{global} $S_2$ and, as described in Appendix~\ref{sec:app:spirals}, avoid the need to trace the crest of a spiral in order to characterize it.

\subsection{Modal Power}
\label{sec:spirals:modal_power}

\begin{figure*}
    \centering
    \includegraphics[]{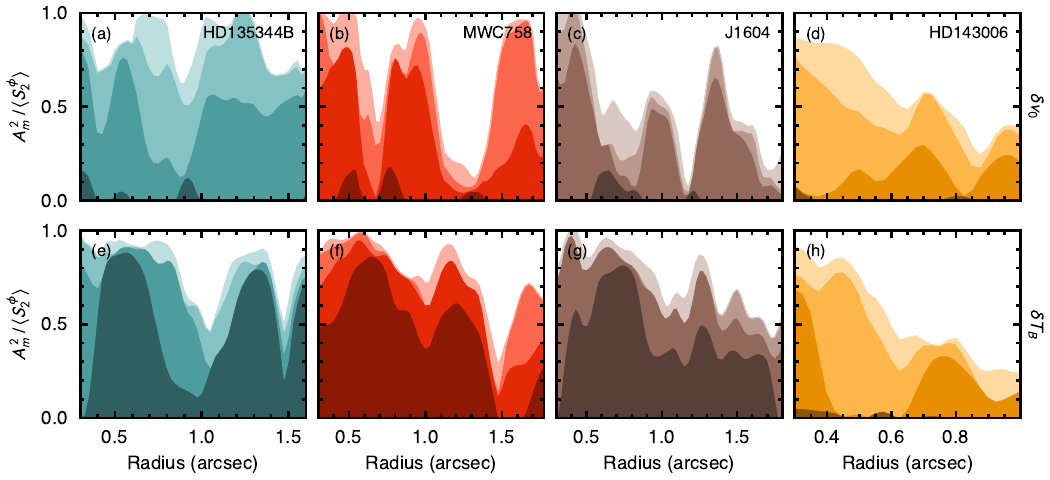}
    \caption{Modal power as a function of radius for each source and both channels. The y-axis shows the fraction of the total azimuthal variance at a given radius that can be accounted for by $m = 1 - 4$ modes. Darker swathes represent lower-order terms. The top row shows the velocity residual channel, while the bottom shows the peak intensity channel. $m=1$ modes appear to be more prevalent in the intensity channel than the velocity as the $m=1$ modes were subtracted as part of the background model as described in Section~\ref{sec:substructures:velocity}.}
    \label{fig:modal_power_radial}
\end{figure*}

The periodicity of the azimuthal structure allows us to differentiate between structured fields, like grand-design spiral arms and shadows, and semi-random fields such as flocculent spirals. The per-annulus azimuthal Fourier amplitude is the standard diagnostic of this distinction in simulations of self-gravitating disks \citep{Cossins_ea_2009, Longarini_ea_2025b}; we measure the same quantity through $S_2$, which tolerates the masked and incompletely covered annuli of a deprojected map and admits a matched noise null. To measure the periodicity, we fit $S_2(\delta \phi)$ at each reference radius, taking the $\delta r = 0$ cut, with the form
\begin{equation}
    S_2(\delta \phi) = N_\phi + \sum_{m} A_m^2 \big(1 - \cos (m\cdot\delta \phi)\big)
    \label{eq:modal_power}
\end{equation}
where $N_\phi$ describes the noise floor (i.e., incoherent variance) and $A_m$ describes the amplitude of the $m$th mode. The $(1 - \cos)$ form is to account for the fact that the structure function considers the difference in values rather than the similarity, as the autocorrelation function would.

We also calculate the concentration, $\mathcal{C}$, of the dominant mode. Each mode's power is normalized by $\langle S_2^\phi \rangle_i$, the mean of $S_2(\delta\phi)$ over all azimuthal lags at annulus $i$, a measure of the total azimuthal variance at that radius. The share of mode $m$ across the disk is the median over annuli,
\begin{equation}
    f_m = \underset{i}{\rm med} \left[ \frac{A_{m,i}^2}{\langle S_2^\phi \rangle_i} \right],
\end{equation}
the dominant mode is the $m$ that maximizes $f_m$, and its concentration is
\begin{equation}
    \mathcal{C} = \frac{f_{m_{\rm dom}}}{\max_{m \neq m_{\rm dom}} f_m},
    \label{eq:concentration}
\end{equation}
the factor by which the dominant mode outweighs its nearest competitor. This parameter quantifies the spectral `purity', or how dominated the field is by a single mode (as a grand-design arm would, $\mathcal{C} > 1$), rather than one where power is spread across multiple modes, as in flocculent or random-phase fields where no mode dominates, $\mathcal{C} \sim 1$, as shown in Appendix~\ref{sec:app:spirals:modes}.

As a measure of significance, we calculate a $z$-score for each $\mathcal{C}$ which describes the difference between the derived concentration with the median of a sample of $N = 1{,}000$ nulls, normalized by the spread in these nulls. We consider $z$-scores greater than 3 to indicate a genuinely concentrated spectrum. We apply this to both fields for all four sources, limiting $m \leq 4$. Figure~\ref{fig:modal_power_radial} shows the amplitude of each of the 4 modes as a fraction of the total variance with darker bands marking lower-order modes. As shown in Fig.~\ref{fig:app:modal_power_ensemble_median}, the typical uncertainty on each of these bands is between 0.01 and 0.1 as determined from the forward modeling of $N=1{,}000$ noise-only models. HD~135344B and MWC~758 show the strongest evidence of periodic structure across both channels, as clearly seen in Fig.~\ref{fig:onsky_residuals}. Table~\ref{tab:spirals} collects the dominant mode, its share of the azimuthal variance, the concentration and its significance for all eight source--field combinations.

\begin{deluxetable}{cccccccc}
\tablewidth{\columnwidth}
\tablecaption{Spiral Characterization: Modal Power and Winding Contrast\label{tab:spirals}}
\tablehead{
    \colhead{} & \colhead{} & \multicolumn{4}{c}{Modal Power} & \multicolumn{2}{c}{Winding Contrast} \\
    \cmidrule(lr){3-6} \cmidrule(lr){7-8}
    \colhead{Source} & \colhead{Channel} & \colhead{$m$} &
    \colhead{$A_{m}^2 / \langle S_2^\phi \rangle$} &
    \colhead{$\mathcal{C}$} & \colhead{$z$} &
    \colhead{$\beta_\star$ ($\degr$)} & \colhead{$\mathcal{R}$}
}
\startdata
\multirow{2}{*}{J1604}      & $\delta v_0$ & 2 & 0.33 &  $\phn4.8$ & $10.9$ & $+34.3^{+17.7}_{-0.8}$ & $\phn4.7$ \\
                            & $\delta T_B$ & 1 & 0.37 &  $\phn1.8$ & $\phn2.1$ & $-51.9^{+4.4}_{-3.5}$ & $\phn2.9$ \\ \hline
\multirow{2}{*}{HD~135344B} & $\delta v_0$ & 2 & 0.52 &  $\phn1.9$ & $\phn1.9$ & $+30.5^{+1.3}_{-0.3}$ & $\phn6.5$ \\
                            & $\delta T_B$ & 1 & 0.56 &  $\phn8.1$ & $23.0$ & $-42.4^{+1.9}_{-3.3}$ & $\phn1.9$ \\ \hline
\multirow{2}{*}{HD~143006}  & $\delta v_0$ & 3 & 0.28 &  $\phn2.0$ & $\phn1.1$ & $-74.5^{+3.2}_{-15.5}$ & $\phn0.7$ \\
                            & $\delta T_B$ & 3 & 0.14 &  $\phn1.1$ & $-1.0$ & $+75.2^{+26.3}_{-14.1}$ & $\phn0.3$ \\ \hline
\multirow{2}{*}{MWC~758}    & $\delta v_0$ & 2 & 0.31 &  $\phn2.1$ & $\phn2.5$ & $+11.6^{+76.5}_{-0.0}$ & $13.0$ \\
                            & $\delta T_B$ & 1 & 0.52 &  $\phn2.7$ & $\phn4.8$ & $+35.9^{+1.4}_{-1.8}$ & $\phn6.9$ \\
\enddata
\tablecomments{Modal power (Section~\ref{sec:spirals:modal_power}): $m$ is the dominant mode, $A_m^2 / \langle S_2^\phi \rangle$ its median share of the azimuthal variance, and $\mathcal{C}$ its concentration with significance, $z$. Winding contrast (Section~\ref{sec:spirals:winding_contrast}): Positive $\beta_\star$ denotes a counter-clockwise winding on the sky, and its uncertainty is the 16th--84th percentile range over $N = 1{,}000$ noise-perturbed scans, while $\mathcal{R}$ characterizes the significance of the winding relative to an unwound field.}
\end{deluxetable}

The modal spectra allow us to identify the disks in which coherent global modes accompany the small-scale structure from those in which they do not. In the temperature residuals $m=1$ dominates in three of the four disks, and is strongly concentrated in HD~135344B and MWC~758, indicative of large-scale asymmetries; J1604 falls short of the threshold, and HD~143006, after its shadow subtraction, is not at all. In the velocity residuals, from which the background fit removes most of the $m=1$ power, $m=2$ is preferred in J1604, HD~135344B and MWC~758, but only J1604 is concentrated. Elsewhere the azimuthal power is spread across several modes, which is what a field of independently sheared perturbations produces.

\subsection{Winding Contrast}
\label{sec:spirals:winding_contrast}

\begin{figure*}
    \centering
    \includegraphics[]{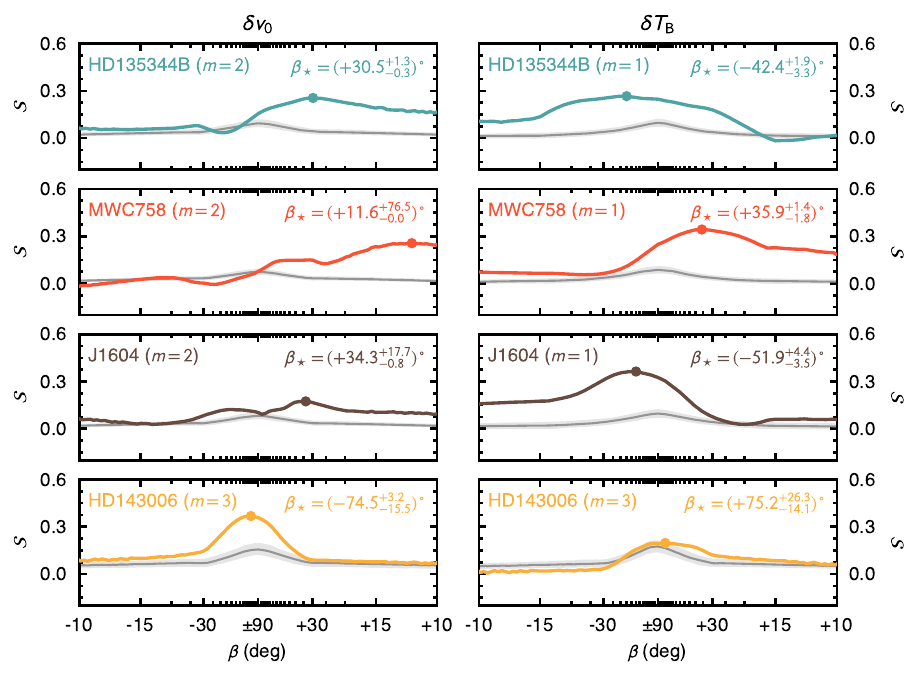}
    \caption{The winding contrast score $\mathcal{S}$ as a function of trial pitch angle $\beta$, for the velocity ($\delta v_0$) and peak brightness temperature ($\delta T_{\rm B}$) residuals of each of the four sources (rows, labeled in panel), in the left and right columns respectively, each scanned under the window of its dominant mode $m$ (labeled in panel; Table~\ref{tab:spirals}). The colored lines show the observations, while the gray line and shaded band show the median and $16$--$84$th percentile range of the $N=1{,}000$ noise-only null ensemble, analyzed identically. The dot marks the peak $\mathcal{S}$, which signifies the preferred $\beta$; $\beta_\star$ and its uncertainty (Table~\ref{tab:spirals}) are labeled in the top right of each panel. The abscissa is linear in the trial slope $s$ but is ticked in $\beta$. The curves show the raw score rather than the excess over the unwound reference at $\beta = \pm 90\degr$, so each retains its own offset and the depth $\Delta\mathcal{S}$ is the height of the marker above its own panel's central value.}
    \label{fig:scan_excess}
\end{figure*}

A measure of modal power informs us of the periodicity of azimuthal variations but it does not provide any phase information, precluding any statement on the possible winding of the structure (radial dependence on the phase). To characterize possible winding, we use a `winding contrast' analysis which quantifies the `tilt' in the global $S_2$. This approach is blind to the periodicity of the structure, but provides a characterization of the mixing of radial and azimuthal correlations across some radial range.

We resample each residual map onto a grid uniform in $u = \ln r$ and azimuth, and compute its global $S_2(\delta u, \delta\phi)$. A logarithmic spiral, $r \propto \exp(\phi \tan\beta)$ with pitch angle $\beta$, will manifest as a linear trough in $S_2$ through the origin, along $\delta\phi = s\,\delta u$, whose gradient $s = (180\degr/\pi)/\tan\beta$ gives the pitch. We score each trial gradient by the mean depth of $S_2$ along that line, each row measured against its own median,
\begin{equation}
    \mathcal{S}(s) = \frac{1}{\Delta S_2\, N_s} \sum_{k=1}^{N_s} \Big[ \underset{\delta\phi}{\rm med}\, S_2(\delta u_k, \delta\phi) - S_2(\delta u_k,\, s\,\delta u_k) \Big],
    \label{eq:score}
\end{equation}
where $|\delta\phi| \leq 180\degr/m$, and $\Delta S_2 = \max S_2 - \min S_2$ (which is used to makes the score dimensionless). Subtracting each row's median removes the radial envelope of $S_2$, so that $\mathcal{S}$ measures only how strongly correlated the field is along the trial line. The preferred gradient maximizes $\mathcal{S}$. Appendix~\ref{sec:app:spirals:winding_contrast} gives the trial range and demonstrates the statistic on fields of known truth.

Figure~\ref{fig:scan_excess} shows the winding contrast scores for all four sources and both channels with positive $\beta$ describing spirals whose azimuth increases with radius --- a counter-clockwise winding on the sky as displayed, north up and east left --- and negative $\beta$ describing the opposite winding. The colored lines show $\mathcal{S}$ derived from the observations, while the gray lines beneath show the null ensemble ($N=1{,}000$ noise-only models analyzed in the same way). The dot marks the peak $\mathcal{S}$ which can be interpreted as corresponding to the preferred pitch angle, $\beta_\star$. The scan requires an azimuthal mode $m$, which sets the window $|\delta\phi| \leq 180\degr/m$ over which trial lines are drawn. For each field we take $m$ to be the dominant mode of its modal decomposition.

We refine $\beta_\star$ below the search-grid spacing with a parabola through the three highest points of the scan, analogous to the quadratic peak fit used to centroid lines in \texttt{bettermoments} \citep{Teague_Foreman-Mackey_2018}. Its uncertainty, $\delta\beta_\star$, is empirically derived. We add each of the $N = 1{,}000$ nulls to the data (increasing the noise amplitude by a factor of $\sqrt{2}$, but not changing the spatial properties of the field), and measure $\beta_{\star}$. We take the 16th--84th percentile range of these values as the uncertainty.

Maximizing $\mathcal{S}$ reveals the preferred winding of the structure, but alone it does not provide us a level of significance. To quantify this, we first take the depth of the peak relative to radial spokes,
\begin{equation}
    \Delta\mathcal{S}(\beta_\star) = \mathcal{S}(\beta_\star) - \mathcal{S}(\pm 90\degr),
    \label{eq:depth}
\end{equation}
which indicates whether some level of winding is preferred over no winding (which is \emph{not} the same as no azimuthal variations). We additionally measure each of the $N = 1{,}000$ null draws in the same way, at its own best slope $\beta_{\star, i}$, and define
\begin{equation}
    \mathcal{R} = \frac{\Delta\mathcal{S}_{\rm obs}(\beta_\star)}
    {{\rm med}_i\big[\Delta\mathcal{S}_{{\rm null}, i}(\beta_{\star, i})\big]},
    \label{eq:winding_R}
\end{equation}
so that $\mathcal{R} \simeq 1$ means the observed winding is no stronger than a pure-noise field produces by chance. Across the eight nulls, only $5\%$ of noise draws reach $\mathcal{R} = 4$ and $1\%$ reach $\mathcal{R} \approx 5$. Every field we describe as wound lies beyond the 99th percentile of its own null, and every other field within the 90th. Table~\ref{tab:spirals} lists $\beta_\star$, with its uncertainty, and $\mathcal{R}$.

We interpret $\beta_\star$ differently for the two fields. Projections of velocity components along the line-of-sight mixes $v_z$ with the projected $v_\phi$ and $v_r$, each of which can carry a differently wound pattern. This is most clearly seen for the case of vortex-driven spirals as considered in \citet{Wolfer_ea_2025}. To avoid misinterpretation, we consider only whether a velocity field is wound, while for the temperature field, which does not suffer from projection effects, we also report the winding direction.

The velocity residuals of J1604, HD~135344B and MWC~758 are significantly wound, with MWC~758 being the most significantly in the sample, while HD~143006 is consistent with noise ($\mathcal{R} < 1$). In the temperature residuals only MWC~758 is wound, with the same direction as its velocity field; those of J1604, HD~135344B and HD~143006 are consistent with noise ($\mathcal{R} \leq 2.9$). Scanning the velocity residuals in three radial bands returns a radially uniform sense only for MWC~758; in J1604 and HD~135344B the bands that carry the signal prefer the opposite sense to the disk-averaged value, as expected if differently wound components dominate at different radii.

\section{Discussion}
\label{sec:discussion}

The heuristics of Section~\ref{sec:structure_function} and the spiral statistics of Section~\ref{sec:spirals} point to a common picture of the correlated structure in these disks: it is radially compact, azimuthally sheared and, in the velocity fields of three of the four disks, wound. In this section we consider these physical traits in turn, before placing the results in the context of the exoALMA sample.

\subsection{Radial Scales and the Pressure Scale Height}
\label{sec:discussion:radial}

Because $\hat{\ell}_r$ is limited by the beam, it sets an upper limit on the radial scale of the non-axisymmetric structure. The natural comparison is the gas pressure scale height, $H = c_s / \Omega$, at or below which the radial correlation lengths of the candidate instabilities are expected to lie \citep{Nelson_ea_2013, Guan_ea_2009, Cossins_ea_2009, Flock_ea_2011, Barraza-Alfaro_ea_2025, Teague_ea_2026}. Both $\delta v_0$ and $\delta T_{\rm B}$ are measured at the $^{12}$CO emission surface, so the relevant scale height is the local one, set by the gas temperature which the optically thick $^{12}$CO line traces. We take the azimuthal median of the peak $^{12}$CO brightness temperature at the average radius of each analysis window, $91$--$115$~au, which ranges over $37$--$46$~K. With a mean molecular weight of $2.37$ and the dynamical masses of Table~\ref{tab:sources}, we find $H \approx 9$--$14$~au ($H/r \approx 0.09$--$0.12$).

The measured $\hat{\ell}_r$, $9$--$16$~au, then lie at $0.7$--$1.6\,H$, and the beam itself corresponds to $1.0$--$1.6\,H$. Correlated non-axisymmetric structure on radial scales larger than ${\sim}1.5\,H$ is therefore excluded in all four disks. This is the regime expected of disk turbulence, although at this resolution it cannot separate the vertical shear, magneto-rotational and gravitational instabilities: identifying the vertical shear instability kinematically requires a beam below ${\sim}10$~au \citep{Barraza-Alfaro_ea_2021}, whereas ours spans $20$--$25$~au. Nor can the data test whether the radial scale tracks $H$.

\subsection{Azimuthal Elongation and Shear}
\label{sec:discussion:shear}

Although the radial scale is only an upper limit, the arc length is resolved, and the structure is consistently longer in azimuth than in radius. This is the expected outcome of Keplerian shear acting on small-scale perturbations, and matches the azimuthally elongated correlation functions of magneto-rotational turbulence in shearing-box simulations \citep{Guan_ea_2009}. A patch of initially isotropic structure advected by a flow of shear rate $q\Omega$, with $q = 3/2$ for Keplerian rotation, keeps its radial width while its azimuthal extent grows, so that after a time $t$ it has $\mathcal{A} = [1 + (q\Omega t)^2]^{1/2}$ \citep{Goldreich_Lynden-Bell_1965}. The anisotropies of $1.6$--$1.8$ measured where the azimuthal power is spread across several modes, in the velocity residuals of HD~135344B and MWC~758 and the temperature residual of J1604, which are themselves lower limits, then correspond to structure that has been sheared for at least $0.13$--$0.16$ of an orbit, i.e., that survives for of order a dynamical time before it is sheared away, or is continually regenerated. The larger anisotropies of the $m=1$-dominated temperature residuals of HD~135344B and MWC~758 partly reflect the global pattern itself, as discussed below. The velocity anisotropies are conservative since the background fit removes the largest-scale azimuthal structure from that channel.

The modal spectra help separate the two possible origins of this elongation. A single coherent low-$m$ pattern is itself elongated in azimuth, so $\hat{\mathcal{A}} > 1$ alone does not distinguish one global mode from many independently sheared perturbations (flocculant spirals). Where the azimuthal power is spread across several modes, as in the velocity residuals of HD~135344B and MWC~758 and the temperature residual of J1604, the elongation must come from many independent features. Where it is concentrated in a single mode, as in the temperature residuals of HD~135344B and MWC~758 at $m=1$, part of the elongation is the global pattern itself, so these anisotropies overstate the shear. HD~143006 is the exception to the picture as a whole. After the removal of its shadow it is isotropic within the noise in both fields and carries no concentrated mode.

\subsection{Winding and Angular Momentum Transport}
\label{sec:discussion:winding}

A sheared perturbation is necessarily also a tilted one. Shear turns the long axis of the patch towards the azimuthal direction, always in the trailing sense. Trailing structure is also able to transport angular momentum outward, whether through the Reynolds stress of correlated $\delta v_r$ and $\delta v_\phi$ or as density waves \citep{Lynden-Bell_Kalnajs_1972}, which turbulent fluctuations excite as the shear swings them from leading to trailing \citep{Heinemann_Papaloizou_2009a, Heinemann_Papaloizou_2009b}. The winding contrast allows us therefore to establish whether the structure leans and, where projection allows, in which sense, providing insight as to possible turbulent angular momentum transport.

Three of the four velocity fields are wound ($\mathcal{R} > 4$): J1604, HD~135344B and MWC~758. As discussed in Section~\ref{sec:spirals:winding_contrast}, their sense is not interpretable on its own. The band-by-band scans of J1604 and HD~135344B, in which the sense reverses between the radii that carry the signal, are what line-of-sight mixing of differently wound $v_\phi$, $v_r$ and $v_z$ components would produce, a difficulty that \citet{Wolfer_ea_2025} also note for the kinematic signatures of vortices.

The temperature residual has no such ambiguity, and MWC~758 is the one disk in which it is wound in a trailing direction. We do not, however, read this as structure that transports angular momentum. The temperature residual of MWC~758 is dominated by a concentrated $m=1$ mode, which we attribute to the illumination of the warp proposed by \citet{Winter_ea_2025}, as discussed in Section~\ref{sec:discussion:exoalma}.

The velocity field of MWC~758 is also wound. As with the other disks its sense is hard to interpret, but inspection of Figure~\ref{fig:polar_residuals} suggests a coherent pattern, and it is the only velocity field whose sense is uniform across the radial bands. What can be said unambiguously is that there is small-scale structure: it is radially compact and azimuthally sheared, and its spectrum is not significantly concentrated. Structure of this kind is what could redistribute angular momentum.

The temperature residuals of J1604 and HD~135344B are no more wound than noise ($\mathcal{R} \leq 2.9$), and so carry no information on the winding sense. HD~143006 is wound in neither field, returning a near-radial preferred slope and $\mathcal{R} < 1$, no more winding than a pure-noise field of the same amplitude finds by accident.

\subsection{In the Context of the exoALMA Sample}
\label{sec:discussion:exoalma}

Before placing these results in the context of previous exoALMA results, it is important to recall that the background modeling applied here differs from the \texttt{discminer} analysis used in other exoALMA manuscripts. The most notable implications of this are the removal of the large, single-armed velocity spiral found in MWC~758 (which \citealt{Winter_ea_2025} argues is the manifestation of a warp), the annular ring in the velocity perturbations of J1604 \citep{Stadler_ea_2023}, and, in the temperature residual, the shadow of HD~143006 \citep{Benisty_ea_2018}, which we subtract explicitly.

Nonetheless, the structure that we recover is consistent with what was previously found. \citet{Fukagawa_ea_2026} identified spiral-like features in MWC~758 and HD~135344B, a ring-like structure in J1604 and a shadow-driven arc in HD~143006, and \citet{Benisty_ea_2026} separated the vertical-flow residuals of these disks into large spirals for MWC~758 and patches or arcs for HD~135344B. Our winding contrast reproduces this picture: MWC~758 is wound coherently across the disk, HD~135344B shows evidence for winding, but multiple components appear to overlap, and HD~143006 has no winding at all. J1604, whose ring our background model removes, retains a wound velocity residual. The analysis here allows us to quantify how significantly this winding deviates from a pure-noise model.

Our results also support the warp interpretation of \citet{Winter_ea_2025}. Our temperature residuals are dominated by $m = 1$ in three of the four disks, at amplitudes $\hat{\sigma} = 2.1-3.9$~K, and in two of those three with no winding beyond what noise produces ($\mathcal{R} \leq 2.9$). This structure is that of a lopsided illumination, consistent with the structure expected from moderate warping \citep{Teague_ea_2022}, and in MWC~758, where the $m=1$ temperature structure is also wound, with the warp that \citet{Winter_ea_2025} propose. In the velocity channel, however, the background fit absorbs the single-armed signature of the warp, and so the $m = 2$ winding we detect in MWC~758 is unlikely to be due to a warped geometry \citep[although the sloshing and breathing motions associated with warps may still be present;][]{Kimmig_Dullemond_2024}. A more complete warp model will be needed to fully interpret these residuals.

\section{Conclusions}
\label{sec:conclusions}

We have applied the second-order structure function framework of \citet{Teague_ea_2026} to the velocity and peak brightness temperature residuals of four near-face-on exoALMA disks, in each case measured against a matched noise-only null built from the data themselves, and have extended it with two statistics that characterize spiral structure without tracing an arm crest. Taken together, the measurements describe the correlated structure of these disks in three ways that inspection of a residual map does not:
\begin{itemize}
    \item The structure is radially compact. Correlated structure is detected in all eight source--field combinations, but its radial correlation length is limited by the $0\farcs15$ beam in every case. This excludes non-axisymmetric structure on radial scales larger than ${\sim}1.5$ times the local pressure scale height at the $^{12}$CO-emitting layer, and is consistent with the sub-scale-height structure expected of disk turbulence.
    \item The structure is sheared. In HD~135344B, MWC~758 and the temperature field of J1604 it is resolved and elongated in azimuth by factors of $1.6$--$2.3$, which are lower limits. Where the azimuthal power is spread across several modes, so that the elongation is not a global pattern, a passively sheared perturbation requires at least $0.13$--$0.16$ of an orbit, so the structure survives for of order a dynamical time. Coherent global modes sit on top of this sheared field only where the modal spectrum is concentrated.
    \item The structure is wound. The velocity fields of three of the four disks are significantly wound, which the structure function detects without tracing a crest, although line-of-sight projection prevents their sense from being interpreted on its own. Only in MWC~758 is the winding coherent across the disk and shared by the temperature field. Its trailing $m=1$ temperature structure is consistent with the illumination of the warp proposed for this disk, while its velocity field shows small-scale, sheared structure that could redistribute angular momentum.
\end{itemize}

We have demonstrated that a statistical analysis of the velocity perturbations offers a new way to characterize the ubiquitously observed structure that we have thus far struggled to characterize parametrically. The application of these techniques to more inclined sources is a clear next step once a method for robustly disentangling emission from the near and far sides of the disk has been developed \citep[e.g.,][]{Izquierdo_ea_2025}.

\begin{acknowledgments}
We thank the anonymous referee for their insightful comments which have considerably improved the clarity and concision of the manuscript.
This paper makes use of the following ALMA data: ADS/JAO.ALMA\#2021.1.01123.L. ALMA is a partnership of ESO (representing its member states), NSF (USA) and NINS (Japan), together with NRC (Canada), MOST and ASIAA (Taiwan), and KASI (Republic of Korea), in cooperation with the Republic of Chile. The Joint ALMA Observatory is operated by ESO, AUI/NRAO and NAOJ. The National Radio Astronomy Observatory is a facility of the National Science Foundation operated under cooperative agreement by Associated Universities, Inc.
This work was supported by NASA's Exoplanets Research Program through grant No. 80NSSC26K1149. 
MB-A acknowledges funding from the European Union's Horizon Europe research and innovation programme under the Marie Skłodowska-Curie grant agreement No. 101281707 (InnerDisk). SF acknowledges financial contribution from the
European Union (ERC, UNVEIL, 101076613). Views and opinions expressed are however those of the author(s) only and do not necessarily reflect those of the European Union, the European Research Executive Agency or the European Research Council. Neither the European Union nor the granting authority can be held responsible for them.
\end{acknowledgments}

\begin{contribution}
RT was responsible for designing the experiment, carrying out the analysis and writing the manuscript.
All authors reviewed the manuscript, helped edit it, and offered suggestions and clarifications.
\end{contribution}

\facilities{ALMA}

\software{astropy \citep{astropy2013, astropy2018, astropy2022},
          bettermoments \citep{Teague_Foreman-Mackey_2018},
          eddy \citep{eddy},
          emcee \citep{emcee},
          GoFish \citep{GoFish},
          matplotlib \citep{matplotlib},
          numpy \citep{numpy},
          scipy \citep{scipy}}
          
\appendix

\section{Observations}
\label{sec:app:observations}

Table~\ref{tab:sources} details the parameters we used for generating the moment maps. Values without an uncertainty were held fixed and taken from \citet{Teague_ea_2025}. Those with an uncertainty are the median posterior values from an \texttt{eddy} fit to the data assuming a pure-Keplerian rotation profile.

Figure~\ref{fig:polar_residuals} shows the polar-projection of the residuals for the four sources, limiting the radial range to that used for the structure function analysis. The top row shows the velocity perturbations while the bottom shows the peak brightness temperature variations. We removed the shadow from HD~143006 by modeling it as a single sinusoidal variation where the amplitude and phase were optimized for at each radius.

\begin{deluxetable*}{l*{10}c}
\tablecaption{Source properties.\label{tab:sources}}
\tablehead{
    \colhead{Source} & \colhead{Rotation} & \colhead{$i$} &
    \colhead{PA} & \colhead{$d$} &
    \colhead{$x_0$} & \colhead{$y_0$} & \colhead{$M_{\rm dyn}$} &
    \colhead{$v_{\rm LSR}$} & \colhead{$r_{\rm in}$} &
    \colhead{$r_{\rm out}$} \\
    \colhead{} & \colhead{} & \colhead{($\degr$)} &
    \colhead{($\degr$)} & \colhead{(pc)} &
    \colhead{(mas)} & \colhead{(mas)} & \colhead{($M_\odot$)} &
    \colhead{(${\rm m\,s^{-1}}$)} & \colhead{(\arcsec)} &
    \colhead{(\arcsec)}
}
\startdata
J1604      & CW  & 6.0     & 258.1 & 145 & $-83.5 \pm 1.0$ & $2.3 \pm 1.0$  & $1.281 \pm 0.003$ & $4621.0 \pm 0.4$ & 0.3 & 1.8 \\
HD~135344B & CCW & $-16.1$ & 242.9 & 135 & $-33.2 \pm 0.4$ & $8.1 \pm 0.4$  & $1.602 \pm 0.002$ & $7086.2 \pm 0.4$ & 0.3 & 1.6 \\
HD~143006  & CW  & $-16.9$ & 168.9 & 167 & $1.7 \pm 0.6$   & $29.2 \pm 0.6$ & $1.323 \pm 0.003$ & $7716.7 \pm 0.8$ & 0.3 & 1.0 \\
MWC~758    & CW  & 19.4    & 240.3 & 156 & $-17.4 \pm 0.4$ & $49.1 \pm 0.5$ & $1.388 \pm 0.002$ & $5885.1 \pm 0.6$ & 0.3 & 1.8 \\
\enddata
\tablecomments{Parameters without uncertainties are held fixed at the
\citet{Teague_ea_2025} values. The sign of $i$ encodes the orientation of the near side rather than the rotation direction: following \citet{Teague_ea_2025}, a negative value places the northern edge of the disk away from the observer. The direction of rotation on the sky --- CW for clockwise, CCW for counter-clockwise --- is an independent property, taken from \citet{Izquierdo_ea_2025}. The source center $(x_0, y_0)$, dynamical mass $M_{\rm dyn}$ and systemic velocity $v_{\rm LSR}$ are fit jointly to the velocity map of Section~\ref{sec:substructures:velocity} with \texttt{eddy}, over $r_{\rm in} \le r \le r_{\rm out}$; the same radial range bounds the structure function analysis of Section~\ref{sec:structure_function}, with $r_{\rm in} = 2\theta_b$. Quoted values are posterior medians and the uncertainties are the statistical $16$th--$84$th percentile half-widths. These are lower limits on the true uncertainties as they do not include systematic terms and underestimate the true statistical uncertainties by a factor of ${\sim}10$ \citep{Hilder_ea_2025}.}
\end{deluxetable*}

\section{Noise Models and Propagation}
\label{sec:app:noise_models}

\begin{figure*}
    \centering
    \includegraphics[]{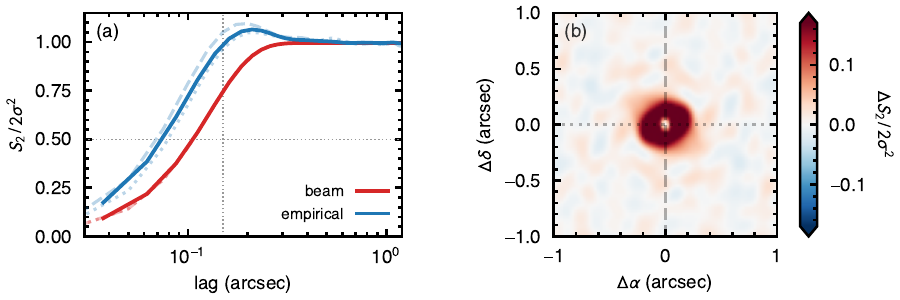}
    \caption{A comparison of an empirical and analytical structure function calculated for noise from the HD~135344B dataset. Panel (a) shows the azimuthally averaged $S_2$ profiles calculated analytically, red, and empirically, blue, each normalized by the large-lag plateau, $2\sigma^2$. The dashed and dotted lines show cuts along $\Delta \alpha$ and $\Delta \delta$ directions, respectively. Panel (b) shows the difference of the two structure functions. The smaller half-power lag $\delta_{1/2}$ of the empirical $S_2$ compared to the analytical $S_2$ suggests that the analytical approach over-predicts the correlation range, while the enhancement in the empirical $S_2$ beyond the large-lag plateau is indicative of anti-correlation in the data; essentially the impact of negative side-lobes.}
    \label{fig:app:noise_comparison}
\end{figure*}

The noise model proposed in \citet{Teague_ea_2026} was spatially correlated white noise with the CLEAN beam of the data being used as the correlation kernel generated in the velocity space. This has three issues. Firstly, such a model is correlated on the wrong scale. Convolving white pixel noise with a beam of FWHM $\theta_b$ yields a field whose autocorrelation is a Gaussian function that is broader by $\sqrt{2}$ than the PSF. Secondly, the clean beam only models the core of the PSF and neglects the (sometimes substantial) side lobes of the dirty beam, which no Gaussian reproduces at any width. That the PSF is genuinely non-Gaussian here is recorded in the images themselves: the Jorsater--van Moorsel ratio of the clean-beam volume to the PSF volume within its first null \citep{Jorsater_vanMoorsel_1995, Czekala_ea_2021} is $\epsilon = 0.55$, $0.74$, $0.63$ and $0.65$ for J1604, HD~135344B, HD~143006 and MWC~758 respectively, so between a quarter and a half of the main-lobe volume lies outside the fitted Gaussian in every case. Applying the JvM correction would not remove either effect: it rescales the residual map by the scalar $\epsilon$, which restores the flux scale but leaves the correlation length untouched. Thirdly, while the noise in the data is typically characterized across several, signal-free channels, we are rather working in a different data space, namely some collapsed statistical view of the data. We must therefore account for this manipulation of the data and propagate our noise model through to the end. In this Appendix we build on the noise model of \citet{Teague_ea_2026} and propose an updated pipeline for accounting for the noise.

\subsection{Empirically Derived $S_2$}
\label{sec:app:noise_models:empirical}

Figure~\ref{fig:app:noise_comparison} shows the difference between an empirically derived $S_2^{\rm noise}$ for line-free channels (using the \texttt{noise\_structure\_function} function in \texttt{eddy}) and an analytical $S_2^{\rm gauss}$ calculated based on the CLEAN beam size and orientation and the variance of those same channels. Panel (a) shows the azimuthally averaged $S_2$ normalized by twice the variance of the noise such that the plateau converges to unity. Panel (b) shows the difference between the two structure functions, again in the same normalization. It is clear that $S_2^{\rm noise}$ crosses half its plateau at a smaller lag than $S_2^{\rm gauss}$, $\delta_{1/2} = 0\farcs075$ against $0\farcs106$, or $\ell = 0\farcs064$ against $0\farcs090$ in the convention of Section~\ref{sec:structure_function:heuristics}: the data are correlated on a scale $\sqrt2\times$ shorter than the CLEAN beam properties predict as described previously. It is therefore this $0\farcs064$, and not the $0\farcs090$ that the beam-convolved model implies, that is the floor against which the radial correlation lengths of Section~\ref{sec:structure_function:results} should be read.

The empirical $S_2^{\rm noise}$ also highlights the impact that the beam side-lobes have on the data. The structure function overshoots its plateau, peaking at $1.065 \times 2\sigma^2$ at a lag of $\approx 0\farcs21$, demonstrating that there is \emph{anti-correlation} in the data on those scales: the imprint of negative side-lobes. Measured against the Gaussian form, the peak fractional excess reaches $0.27$ of the plateau azimuthally averaged and $0.34$ in the full two-dimensional $S_2$.

\begin{figure*}
    \includegraphics[width=0.96\textwidth]{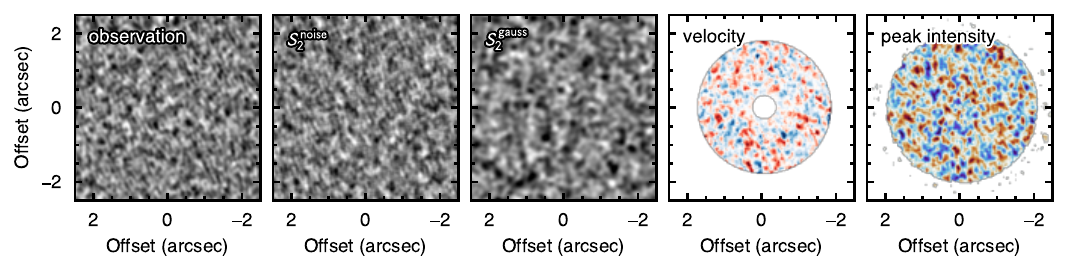}
    \caption{Comparisons of the different noise models, here for J1604. The left panel shows the structure of a noise-only channel map from the observation. The second and third panels show a noise draw based on the empirically determined $S_2^{\rm noise}$ and analytical $S_2^{\rm gauss}$, respectively, both matched in amplitude to the data. The analytical form clearly over-estimates the spatial correlations in the data, as shown in Fig.~\ref{fig:app:noise_comparison}. The right two panels show the empirical noise model propagated through to the velocity map residuals and peak intensity residuals, respectively, including the same clipping and smoothing applied to the data. All panels but the fourth are in ${\rm mJy\,beam^{-1}}$, the fourth in ${\rm m\,s^{-1}}$; each is scaled independently to $\pm 3$ times its own rms so that the morphologies, rather than the amplitudes, are compared.}
    \label{fig:app:noise_examples}
\end{figure*}

The difference between the two noise models is most clearly seen in the left three panels of Figure~\ref{fig:app:noise_examples} which plot a line-free channel from the J1604 data, a draw from $S_2^{\rm noise}$ and a draw from $S_2^{\rm gauss}$, respectively. The $S_2^{\rm gauss}$ draw appears to lack the fine-scale structure of both the true data and the $S_2^{\rm noise}$ characterization, despite being matched in amplitude. We strongly recommend, therefore, that noise models are based on empirical measurements if possible, rather than just relying on the CLEAN beam properties.

\subsection{Propagation to Moment Maps}
\label{sec:app:noise_models:propagation}

Once noise has been characterized in the image plane, it is possible to pass that through the entire analysis pipeline to build a suite of null samples. The right two panels of Figure~\ref{fig:app:noise_examples} show the resulting null model for the velocity and peak intensity, respectively.

There is a good reason to do this at the level of the cube rather than at the level of the residual map. The uncertainty on a line centroid scales roughly as the line width divided by the peak signal-to-noise ratio, so as $F_\nu$ and $\Delta V$ vary with radius the velocity noise becomes strongly heteroscedastic even though the image-plane noise it derives from is not. A single, stationary structure function is therefore an adequate description of the noise in the channel maps, where we measure it, but not of the noise in the velocity or peak brightness temperature residuals, where we need it. Propagating realizations through the collapse recovers this radial dependence for free.

There is no clear preference for structure in the velocity or peak intensity residuals, however we do find a large source-to-source variation in the null ensembles and between velocity and peak intensity. We find that the noise floor varies by about a factor of 1.8 for the velocity residuals across the four sources, but only by a factor of 1.4 for the peak intensity. The difference between the velocity and intensity noise floors is due to the radial heteroskedasticity (non-uniformity of the errors) of the centroiding method described above, expressing itself as a source-to-source difference in brightness and line width.

Equally important, the observation and the null are passed through the \emph{identical} extraction. The biases inherent in the measurement --- principally the sensitivity of the centroid to intensity gradients across the beam --- are then common to both and cancel in the comparison, rather than needing to be modeled. The polar-coordinate representation of Section~\ref{sec:structure_function:stacks} needs no such cancellation, being absorbed by the choice of convention rather than left in the measurement \citep{Teague_ea_2026}. For the same reason, we retain the ensemble of per-draw structure functions as the null instead of averaging them into a single $S_2$. Collapsing and re-synthesizing would reimpose stationarity and discard exactly the radial structure we have gone to the trouble of capturing.

As we are working with sources that have strong velocity and intensity gradients, it is important to include the background line emission as part of the noise propagation which can be subtly correlated due to these gradients. These are constructed by deriving an azimuthally averaged $F_\nu(r)$ profile, based on the peak intensity moment map, and a similar azimuthally averaged $\Delta V(r)$ profile derived from the \texttt{percentiles} method used to obtain the line center. With a background model of the source to hand, we constructed our noise ensemble with the following steps:
\begin{enumerate}
    \item Empirically measure the noise based on the first and last 10 channels in the data to build $S_2^{\rm noise}$.
    \item Draw a noise-only cube from $S_2^{\rm noise}$.
    \item Add the background disk model assuming a purely Keplerian disk.
    \item Calculate a moment map from the model plus noise using the same smoothing and clipping parameters as the real data.
    \item Perform the background subtraction using the same azimuthal decomposition applied to the data: \texttt{fit\_annuli} for the velocity map, or the azimuthally averaged radial profile for the peak intensity.
\end{enumerate}
We repeat this $N = 1{,}000$ times for each source and each property. Structure functions are measured over $2\theta_b \le r \le r_{\rm out}$, where $r_{\rm out}$ ranges from $1\farcs0$ to $1\farcs8$ depending on the source, and radial lags are capped at $0\farcs5$ so that the plateau is always estimated from lags which the data sample fully (i.e., $\ell_{\rm max} = (r_{\rm out} - r_{\rm in}) / 2$). The velocity and peak brightness temperature maps are clipped at $4\sigma$ and $3\sigma$ respectively, and each null inherits the value used for its own channel. The shadow removal of Section~\ref{sec:substructures:intensity} is applied inside the loop for HD~143006, so that each draw is refitted and subtracted exactly as the observation is; what the comparison holds fixed between a source and its null is the extraction, not the filtering, and a residual is therefore never compared against a null filtered differently from itself.

\subsection{Consequences for the Inferred Structure}
\label{sec:app:noise_models:consequences}

\begin{figure}
    \centering
    \includegraphics[]{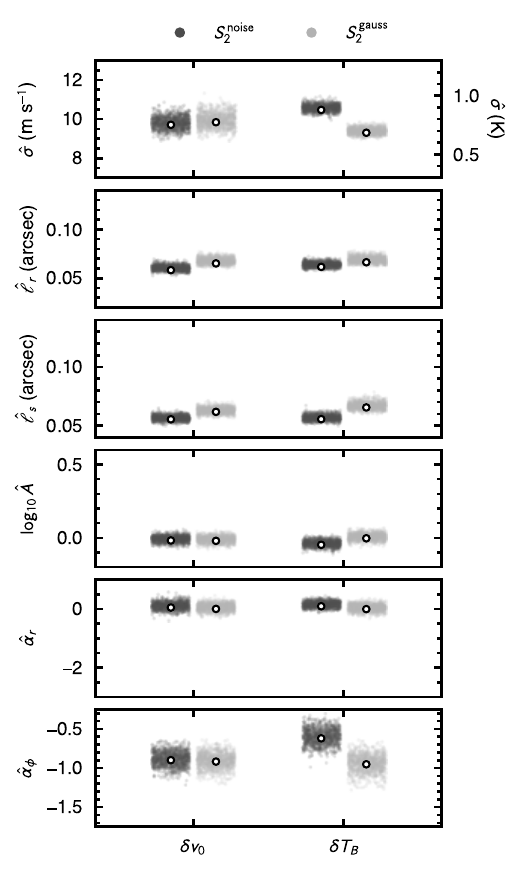}
    \caption{Comparison of the null clouds for the six heuristics and two moment maps for HD~135344B. The dark and light gray points represent the different noise models: $S_2^{\rm noise}$ and $S_2^{\rm gauss}$, respectively, the latter with its kernel at $\theta_b/\sqrt{2}$. The white dots show the median value of each point cloud.}
    \label{fig:app:null_cloud_comparison}
\end{figure}

We compare the empirical null against the beam-convolved model with its kernel set to $\theta_b/\sqrt{2}$, so that, as in \citet{Teague_ea_2026}, the analytic noise is correlated on the CLEAN beam itself rather than on $\sqrt{2}$ times it (Fig.~\ref{fig:app:null_cloud_comparison}). With that correction the analytic null reproduces the empirical length scales to ${\sim}10\%$, the arc length of HD~143006's temperature residual excepted. Repeating the comparison of Section~\ref{sec:structure_function} against it, $\hat{\ell}_r$ and $\hat{\ell}_s$ keep their sign in all eight (source, property) combinations, at a somewhat lower significance. Without the correction the over-wide null places its own $\hat{\ell}_r$ above the observed value, and $\hat{\ell}_r$ changes sign in four of the eight. What the correction does not fix is the peak brightness temperature. Its analytic null is stationary and, for HD~143006, close to isotropic, whereas the empirical nulls are not (see below), and its amplitude is $20$--$35\%$ too low. Against the analytic null the significance of $\hat{\sigma}$ in that channel is therefore inflated, and the azimuthal stationarity slopes lose their meaning: HD~135344B's changes sign, and the steep slopes of MWC~758 and HD~143006 fall below $2\sigma$. A correctly scaled beam model is thus an adequate null for the velocity residuals, but the peak brightness temperature requires the empirical one.

The velocity nulls are close to isotropic and stationary, as the noise model of \citet{Teague_ea_2026} assumes: $\hat{\mathcal{A}} = 0.92$--$1.00$, $\hat{\alpha}_r = -0.17$ to $+0.05$ and $\hat{\alpha}_\phi = -0.90$ to $-1.04$. The peak brightness temperature nulls are not. Their $\hat{\alpha}_\phi$ lies between $-0.88$ and $-0.45$ rather than at the $-1$ of a stationary field, and the HD~143006 null is radially elongated, $\hat{\mathcal{A}} = 0.74$. The beam-convolved noise model does not reproduce either effect (Appendix~\ref{sec:app:noise_models:consequences}), so the peak-intensity extraction itself imposes this structure on the noise.

Two limitations of the empirical model are worth stating. The draws are stationary Gaussian realizations of the measured two-point function, so while they reproduce the correlation structure of the imaged noise, they will not reproduce non-Gaussian imaging artifacts. The channels of each noise cube are also drawn independently, which neglects the channel-to-channel correlation introduced by Hanning smoothing and by regridding to $100~{\rm m\,s^{-1}}$. This is the least justified of our assumptions, given that the moment maps are calculated with a five-channel spectral filter, and it is the natural place to improve the model.

\section{Spirals}
\label{sec:app:spirals}

Section~\ref{sec:spirals} uses two statistics, both built on the $S_2$ machinery of Section~\ref{sec:structure_function}: the \emph{modal power}, Section~\ref{sec:spirals:modal_power}, asks how concentrated it is in a single azimuthal mode, and the \emph{winding contrast}, Section~\ref{sec:spirals:winding_contrast}, asks whether the structure is wound. In this Appendix we provide more details on how these are calculated, provide some intuition for how different phenomenological models behave, and discuss the impact of noise on the measurements.

\subsection{Modal Power}
\label{sec:app:spirals:modes}

A single azimuthal mode of amplitude $A_m$ contributes $A_m^2 (1 - \cos m \delta\phi)$ to the azimuthal $S_2$, which gives a mode decomposition directly from the $S_2$ stacks of Section~\ref{sec:structure_function}. We perform a least-squares fit of Eqn.~\ref{eq:modal_power} to each annulus's azimuthal $S_2$ for $m = 1 - 4$.

\begin{figure*}
    \centering
    \includegraphics[width=\textwidth]{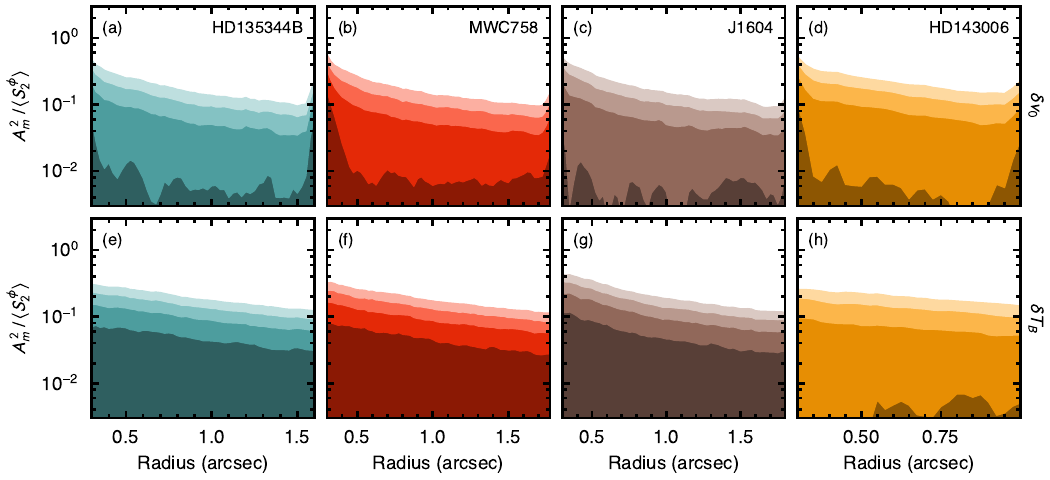}
    \caption{As Figure~\ref{fig:modal_power_radial}, but showing the null ensemble median for each mode, $m$. The ensemble scatter, not shown, is comparable to the median value. Lower-order $m$ modes are more dominant in the peak brightness than the velocity channel.}
    \label{fig:app:modal_power_ensemble_median}
\end{figure*}

\begin{figure*}
    \centering
    \includegraphics[width=\textwidth]{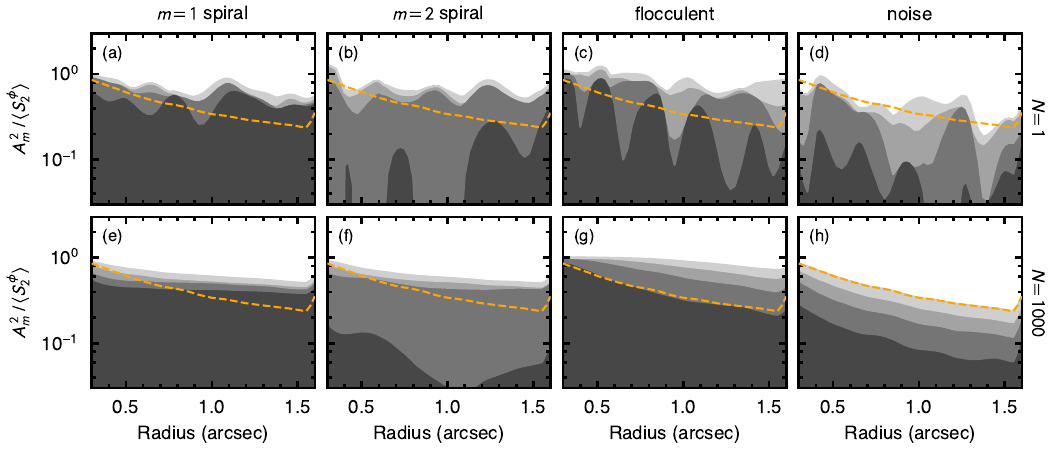}
    \caption{As Figure~\ref{fig:modal_power_radial}, but showing the structure of four phenomenological models of known truth, put through HD~135344B's own deprojection, annulus grid and footprint mask: a coherent $m=1$ arm, a coherent $m=2$ arm, a flocculent field of random azimuthal phase and pure noise, left to right. All four are drawn at ${\rm SNR} \sim 1$. Panels (a)--(d) show a single realization, which is what a disk is; panels (e)--(h) the pair-count weighted combination of $N = 1{,}000$ realizations. The orange dashed line is the low-$m$ total of the noise column's combined stack, repeated in both rows as the floor every correlated field puts into low $m$.}
    \label{fig:modal_power_model_ensembles}
\end{figure*}

Figures~\ref{fig:app:modal_power_ensemble_median} and \ref{fig:modal_power_model_ensembles} calibrate that fit. The first shows what the same statistic returns on fields with no arms in them, offering a representative noise floor for those in Figure~\ref{fig:modal_power_radial} (note the log-scaling used here but not in the main body). The second puts four fields of known truth --- a coherent $m=1$ arm, a coherent $m=2$ arm, a flocculent field carrying the same correlation lengths but random azimuthal phase, and the bare noise --- through the analysis pipeline demonstrating how those morphologies manifest in this projection.

These demonstrate that both coherent arms are recovered at the right amplitude, however the `spectral purity' of the signal, $\mathcal{C}$, is diluted. Measured on the clean polar grid the fields were drawn on, both arms give $\mathcal{C} \approx 9$; measured through the disk's extraction the median falls to $5.2$ and $4.6$ while the floor is unmoved at $1.6$ for the flocculent field and $1.3$ for noise. Since a single disk is a single realization, the relevant width is the spread over draws rather than the ensemble value: the two arms span $3.5 - 7.3$ and $3.2 - 6.9$ at the 16th--84th percentile, against $1.2 - 2.8$ and $1.1 - 1.7$ for the two incoherent fields. A coherent arm at this signal-to-noise therefore delivers $\mathcal{C} \approx 5$ rather than $\approx 9$.

Panel (g) of Figure~\ref{fig:modal_power_model_ensembles} is the reason we quote $\mathcal{C}$ and not the dominant mode. The flocculent column puts $0.35$ of its azimuthal variance at $m=1$, more than either coherent arm. Read by which mode wins, or by how much of the azimuthal variance a low-order description accounts for at all, a random-phase field appears the most convincing one-armed spiral in the figure. Conversely, when read by concentration, it sits with the noise. Its spectrum is a monotone decline across $m$ rather than a peak, and, in a single realization, its dominant mode moves from annulus to annulus (panel c) where a coherent arm's does not. The first of these is the signature by which flocculent structure is identified in simulations of self-gravitating disks, a spectrum that declines across $m$ rather than peaking at a single mode being what separates cooling-driven from coherently driven spirals there \citep{Cossins_ea_2009, Longarini_ea_2025b}. The second has no direct counterpart in those works, whose corresponding diagnostic is the transience of individual arms, which a single epoch cannot access.

The same distinction separates driving mechanisms in simulations of self-gravitating disks: \citet{Longarini_ea_2025b} find that infall- and cooling-driven instabilities share the same dominant mode, which is set by the disk-to-star mass ratio, and are separated instead by whether the power is concentrated at that mode or spread across the higher ones \citep[see also][]{Cossins_ea_2009}. We cap the decomposition at $m \leq 4$. This cap reflects what the beam and projection leave recoverable rather than the range of modes a disk may carry: a cooling-dominated self-gravitating disk at $q \lesssim 0.1$ would place its dominant mode near $m \approx 5$ and retain appreciable power above it \citep{Cossins_ea_2009}, and such a field would register here as an absence of concentration rather than as an identified mode.

The modal spectra also bear on the mechanism driving the structure, although less decisively than the mode numbers alone suggest. Every source--field combination in the sample favors $m \leq 3$, and where the spectrum is concentrated it is concentrated at low order. For a self-regulated, cooling-dominated gravitational instability the expectation is the opposite: a dominant mode near $m \approx 5$ at $q \approx 0.1$, rising to a mean of $m \approx 10 - 15$ at lower disk masses, with power spread across the higher modes \citep{Cossins_ea_2009}. Concentration into $m \leq 2$ would instead require a disk-to-star mass ratio of $q \gtrsim 0.2$ or a coherent driver such as infall \citep{Longarini_ea_2025b}, only the latter of which may tentatively be possible for MWC~758. We have repeated the analysis described in Section~\ref{sec:spirals:modal_power} but including up to $m=6$ terms, but found no difference in the dominant modes.

Low order alone, however, does not identify the driver. Spiral density waves excited by MRI turbulence occupy a comparable range. In the global stratified simulations of \citet{Flock_ea_2011} the midplane radial velocity peaks at $m = 3 - 5$, with the strongest large-scale density disturbances at $m = 5$. That range sits at the edge of the window we can access given the decomposition is capped at $m \leq 4$. What separates the two pictures here is therefore the concentration rather than the mode number. A turbulent wave field is broad-band and transient, and should register as the flocculent case above, i.e., a spectrum declining across $m$ with no stable dominant mode. The concentrated spectra we measure are not of that form and do require a coherent driver, whether a companion or shadowing; the marginal ones do not exclude a broad-band, internally driven wave field. The comparison is not exact in either direction, however: those works decompose the midplane surface density or midplane velocity, whereas we decompose a projected velocity or temperature residual at the emission surface, and the mapping between a density mode and its kinematic counterpart depends on the amplitude of the spiral and on the disk geometry \citep{Longarini_ea_2021, Terry_ea_2022}.

\subsection{Winding Contrast}
\label{sec:app:spirals:winding_contrast}

\begin{figure*}
    \centering
    \includegraphics[width=\textwidth]{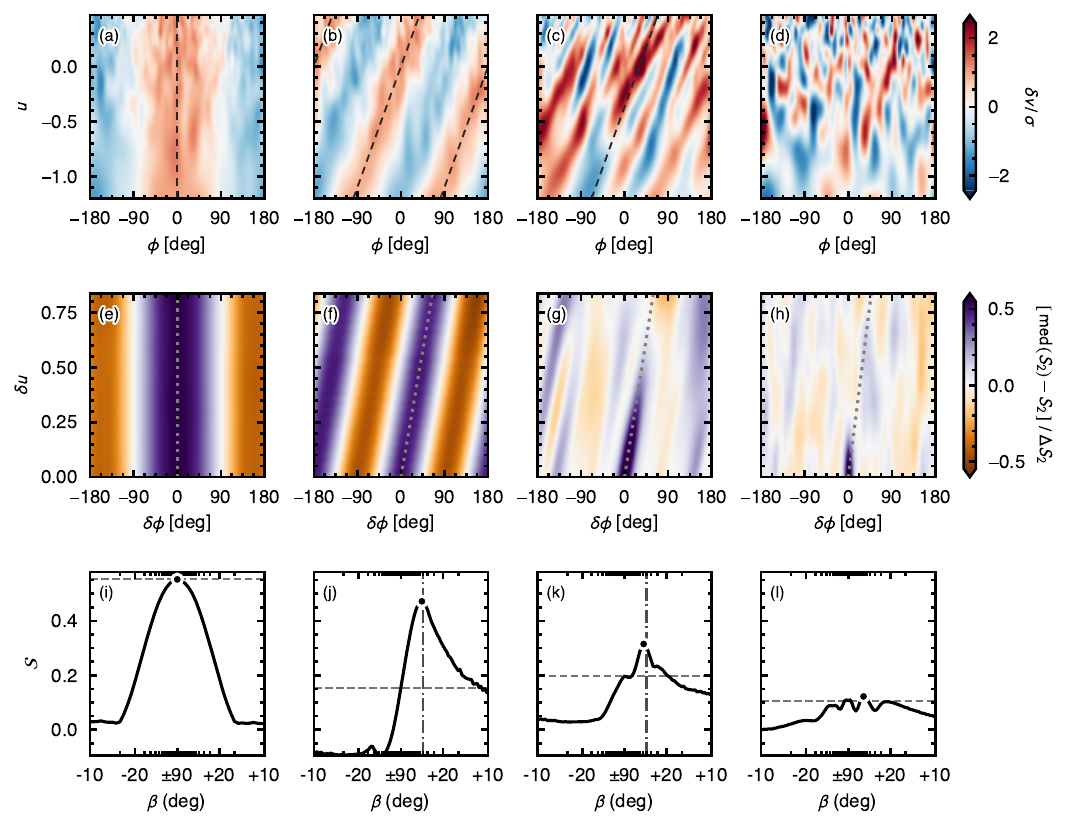}
    \caption{What the winding contrast measures, shown on four synthetic fields of known truth at ${\rm SNR} = 5$ (one shared Gaussian random field, GRF, realization throughout). Columns, left to right: an \emph{unwound} pattern or a radial spoke at $\beta = \pm 90\degr$; a coherent two-armed logarithmic spiral of pitch $\beta = +35\degr$; a \emph{flocculent} field, an anisotropic GRF of random azimuthal phase carrying the same $+35\degr$ lean but no global coherence; and the bare isotropic GRF, i.e.\ the null. The scan is held at $m=1$ in the first column and at $m=2$, the mode of the injected arm, in the other three. Top row (a--d): each field on the log-radial grid, $u = \ln r$ against $\phi$, with dashed lines marking the injected locus where there is one. Middle row (e--h): the surface the scan integrates, normalized by the range of the surface, $[{\rm med}(S_2) - S_2] \, / \, \Delta S_2$, so that a correlated ridge reads as a positive (dark) trough. The dotted line is the recovered slope. Bottom row (i--l): the score $\mathcal{S}$ along that scan, on a slope-linear axis \emph{ticked} in pitch. The filled circle marks the peak $\mathcal{S}(\beta_\star)$, the horizontal dashed line the unwound reference $\mathcal{S}(\pm 90\degr)$ whose difference from the peak is $\Delta\mathcal{S}$, and the vertical dash-dotted line the injected pitch where one exists. Note that the spoke returns the largest $\mathcal{S}$ of the four and yet $\Delta\mathcal{S} \simeq 0$, since its peak is the unwound hypothesis, and that the flocculent field recovers the correct lean at roughly a third of the coherent arm's depth.}
    \label{fig:app:scan_excess_demo}
\end{figure*}

The score of Equation~\ref{eq:score} is evaluated on the global $S_2$ of the residual map which is uniform in $u = \ln r$ over $r_{\rm in} \le r \le r_{\rm out}$ and in azimuth. A coherent $m$-armed logarithmic spiral of pitch $\beta$ produces a trough along $\delta\phi = s\,\delta u$, repeated every $360\degr/m$ in $\delta\phi$, as can be seen in the second row of Fig.~\ref{fig:app:scan_excess_demo}. Restricting the scan to $|\delta\phi| \leq 180\degr/m$, half the period of the pattern, isolates the trough through the origin. The mode $m$ therefore sets both the window within which the per-row median is taken and the extent of the trial lines, and we set it to the dominant mode of each field's modal decomposition (Section~\ref{sec:spirals:modal_power}).

Rows are scored out to $\delta u_{\max} = (\ln r_{\rm out} - \ln r_{\rm in})/2$, half the log-radial baseline, so that every lag is sampled across at least half the disk. We scan $241$ trial gradients uniformly over
\begin{equation}
    |s| \le s_{\max} = \frac{180\degr/m}{\delta u_8},
\end{equation}
where $\delta u_8$ is the eighth row of the grid, so that the steepest line still crosses eight rows before leaving the window; a gradient whose line crosses fewer than eight rows is not scored, as its mean would contain too few points. The odd number of trials samples the unwound hypothesis, $s = 0$, exactly.

To test whether the winding is uniform with radius, we also scan three radial bands, each one third of the log-radial baseline, by applying the same procedure to the corresponding rows of the resampled map, with $\delta u_{\max}$ set to half the band's own baseline and the band's own matched null.

The optimal $\beta$ is determined by the $s$ value which maximizes $\mathcal{S}$, marked as a dot in each panel of the bottom row of Fig.~\ref{fig:app:scan_excess_demo}. This value alone is insufficient as a robust measure of `woundedness'.

Figure~\ref{fig:app:scan_excess_demo} shows why, on four fields of known truth. The first column is a radial spoke, suggestive of a stationary shadow, and it returns the largest peak score of the four while its $\Delta\mathcal{S}$ is identically zero, because the slope it prefers \emph{is} the unwound hypothesis. The second is a coherent two-armed logarithmic spiral of pitch $+35\degr$, recovered at $+36\degr$ and at a depth an order of magnitude above the noise column's. Where an arm is present the scan returns both its sense and its tightness. The fourth is the bare noise field. Its peak falls at $+45\degr$, but at a depth an order of magnitude below the coherent arm's, so a preferred slope means nothing unless its depth is significant. The beam-correlated noise of our nulls favours the unwound hypothesis more strongly still, with its best slope at or next to $\pm 90\degr$ in most draws.

The third column is the limitation the statistic cannot escape. It is an anisotropic Gaussian random field carrying the same $+35\degr$ lean as the coherent arm but with random azimuthal phase (flocculent rather than grand-design) and the scan recovers that lean correctly, at $+39\degr$, with roughly a third of the coherent arm's depth and still several $\sigma$ clear of the noise-only null. A significant $\Delta\mathcal{S}$ therefore establishes that the correlated structure leans, and in which sense, but not that it is a single coherent arm. Scored against a lean-matched flocculent null rather than a noise-only one, the injected arm differs by less than $2\sigma$. This is a property of the estimator and not of the data, and it is why the winding contrast and the modal power are reported together. The two need not agree, and in MWC~758 they do not: it is the most coherently wound source in the sample while its velocity spectrum is not significantly concentrated.

Two practical limits follow. The scan can only express pitch angles down to the tightest slope the log-radial baseline supports, which across our radial windows is ${\approx}2\degr$ at $m=1$ and ${\approx}4\degr$ at $m=2$ and $3$; a peak pinned against that bound would be a limit rather than a measurement, and none of the values in Table~\ref{tab:spirals} approaches it. Taking the difference against $s = 0$ is what makes this a \emph{winding} statistic. Beam-correlated noise is smooth on the log-radial grid and scores a substantial ${\rm depth}$ of its own, but it does so at $s \approx 0$; only the excess over the unwound hypothesis is specific to a spiral. The maximization over slopes costs nothing in the comparison, since every null draw maximizes identically.

While $\beta_\star$ and its interval $\delta\beta_\star$ use the parabola-refined peak described in Section~\ref{sec:spirals:winding_contrast}, $\Delta\mathcal{S}$ and $\mathcal{R}$ are evaluated at the grid peak. The interval measures how precisely the peak is located in the data at hand rather than whether the field is wound: the temperature residual of HD~135344B is located to within a few degrees but has $\mathcal{R} = 1.9$.

\bibliography{main}{}

@ARTICLE{Kimmig_Dullemond_2024,
       author = {{Kimmig}, C.~N. and {Dullemond}, C.~P.},
        title = "{Warped disk evolution in grid-based simulations}",
      journal = {\aap},
         year = 2024,
        month = sep,
       volume = {689},
          eid = {A45},
        pages = {A45},
          doi = {10.1051/0004-6361/202348660},
archivePrefix = {arXiv},
       eprint = {2406.02754},
 primaryClass = {astro-ph.SR},
       adsurl = {https://ui.adsabs.harvard.edu/abs/2024A&A...689A..45K}
}

@ARTICLE{Terry_ea_2022,
       author = {{Terry}, J.~P. and {Hall}, C. and {Longarini}, C. and {Lodato}, G. and {Toci}, C. and {Veronesi}, B. and {Paneque-Carre{\~n}o}, T. and {Pinte}, C.},
        title = "{Constraining protoplanetary disc mass using the GI wiggle}",
      journal = {\mnras},
         year = 2022,
        month = feb,
       volume = {510},
       number = {2},
        pages = {1671-1679},
          doi = {10.1093/mnras/stab3513},
archivePrefix = {arXiv},
       eprint = {2108.11282},
 primaryClass = {astro-ph.EP},
       adsurl = {https://ui.adsabs.harvard.edu/abs/2022MNRAS.510.1671T}
}

@ARTICLE{Longarini_ea_2021,
       author = {{Longarini}, Cristiano and {Lodato}, Giuseppe and {Toci}, Claudia and {Veronesi}, Benedetta and {Hall}, Cassandra and {Dong}, Ruobing and {Patrick Terry}, Jason},
        title = "{Investigating Protoplanetary Disk Cooling through Kinematics: Analytical GI Wiggle}",
      journal = {\apjl},
         year = 2021,
        month = oct,
       volume = {920},
       number = {2},
          eid = {L41},
        pages = {L41},
          doi = {10.3847/2041-8213/ac2df6},
archivePrefix = {arXiv},
       eprint = {2108.11387},
 primaryClass = {astro-ph.EP},
       adsurl = {https://ui.adsabs.harvard.edu/abs/2021ApJ...920L..41L}
}

@ARTICLE{Longarini_ea_2025b,
       author = {{Longarini}, Cristiano and {Price}, Daniel J. and {Kratter}, Kaitlin M. and {Lodato}, Giuseppe and {Clarke}, Cathie J.},
        title = "{Infall-driven gravitational instability in accretion discs}",
      journal = {\mnras},
         year = 2025,
        month = aug,
       volume = {541},
       number = {2},
        pages = {1145-1163},
          doi = {10.1093/mnras/staf1018},
archivePrefix = {arXiv},
       eprint = {2506.13701},
 primaryClass = {astro-ph.EP},
       adsurl = {https://ui.adsabs.harvard.edu/abs/2025MNRAS.541.1145L}
}

@ARTICLE{Cossins_ea_2009,
       author = {{Cossins}, Peter and {Lodato}, Giuseppe and {Clarke}, C.~J.},
        title = "{Characterizing the gravitational instability in cooling accretion discs}",
      journal = {\mnras},
         year = 2009,
        month = mar,
       volume = {393},
       number = {4},
        pages = {1157-1173},
          doi = {10.1111/j.1365-2966.2008.14275.x},
archivePrefix = {arXiv},
       eprint = {0811.3629},
 primaryClass = {astro-ph},
       adsurl = {https://ui.adsabs.harvard.edu/abs/2009MNRAS.393.1157C}
}

@ARTICLE{Loomis_ea_2025,
       author = {{Loomis}, Ryan A. and {Facchini}, Stefano and {Benisty}, Myriam and {Curone}, Pietro and {Ilee}, John D. and {Cataldi}, Gianni and {Yen}, Hsi-Wei and {Teague}, Richard and {Pinte}, Christophe and {Huang}, Jane and {Garg}, Himanshi and {Orihara}, Ryuta and {Czekala}, Ian and {Zawadzki}, Brianna and {Andrews}, Sean M. and {Wilner}, David J. and {Bae}, Jaehan and {Barraza-Alfaro}, Marcelo and {Fasano}, Daniele and {Flock}, Mario and {Fukagawa}, Misato and {Galloway-Sprietsma}, Maria and {Izquierdo}, Andr{\'e}s F. and {Kanagawa}, Kazuhiro and {Lesur}, Geoffroy and {Longarini}, Cristiano and {Menard}, Francois and {Price}, Daniel J. and {Rosotti}, Giovanni and {Stadler}, Jochen and {Wafflard-Fernandez}, Gaylor and {W{\"o}lfer}, Lisa and {Yoshida}, Tomohiro C.},
        title = "{exoALMA. II. Data Calibration and Imaging Pipeline}",
      journal = {\apjl},
         year = 2025,
        month = may,
       volume = {984},
       number = {1},
          eid = {L7},
        pages = {L7},
          doi = {10.3847/2041-8213/adc43a},
archivePrefix = {arXiv},
       eprint = {2504.19870},
 primaryClass = {astro-ph.EP},
       adsurl = {https://ui.adsabs.harvard.edu/abs/2025ApJ...984L...7L}
}

@ARTICLE{Teague_ea_2021,
       author = {{Teague}, Richard and {Bae}, Jaehan and {Aikawa}, Yuri and {Andrews}, Sean M. and {Bergin}, Edwin A. and {Bergner}, Jennifer B. and {Boehler}, Yann and {Booth}, Alice S. and {Bosman}, Arthur D. and {Cataldi}, Gianni and {Czekala}, Ian and {Guzm{\'a}n}, Viviana V. and {Huang}, Jane and {Ilee}, John D. and {Law}, Charles J. and {Le Gal}, Romane and {Long}, Feng and {Loomis}, Ryan A. and {M{\'e}nard}, Fran{\c{c}}ois and {{\"O}berg}, Karin I. and {P{\'e}rez}, Laura M. and {Schwarz}, Kamber R. and {Sierra}, Anibal and {Walsh}, Catherine and {Wilner}, David J. and {Yamato}, Yoshihide and {Zhang}, Ke},
        title = "{Molecules with ALMA at Planet-forming Scales (MAPS). XVIII. Kinematic Substructures in the Disks of HD 163296 and MWC 480}",
      journal = {\apjs},
         year = 2021,
        month = nov,
       volume = {257},
       number = {1},
          eid = {18},
        pages = {18},
          doi = {10.3847/1538-4365/ac1438},
archivePrefix = {arXiv},
       eprint = {2109.06218},
 primaryClass = {astro-ph.EP},
       adsurl = {https://ui.adsabs.harvard.edu/abs/2021ApJS..257...18T}
}

@ARTICLE{Galloway-Sprietsma_ea_2025,
       author = {{Galloway-Sprietsma}, Maria and {Bae}, Jaehan and {Izquierdo}, Andr{\'e}s F. and {Stadler}, Jochen and {Longarini}, Cristiano and {Teague}, Richard and {Andrews}, Sean M. and {Winter}, Andrew J. and {Benisty}, Myriam and {Facchini}, Stefano and {Rosotti}, Giovanni and {Zawadzki}, Brianna and {Pinte}, Christophe and {Fasano}, Daniele and {Barraza-Alfaro}, Marcelo and {Cataldi}, Gianni and {Cuello}, Nicol{\'a}s and {Curone}, Pietro and {Czekala}, Ian and {Flock}, Mario and {Fukagawa}, Misato and {Gardner}, Charles H. and {Garg}, Himanshi and {Hall}, Cassandra and {Huang}, Jane and {Ilee}, John D. and {Kanagawa}, Kazuhiro and {Lesur}, Geoffroy and {Lodato}, Giuseppe and {Loomis}, Ryan A. and {Menard}, Francois and {Orihara}, Ryuta and {Price}, Daniel J. and {Wafflard-Fernandez}, Gaylor and {Wilner}, David J. and {W{\"o}lfer}, Lisa and {Yen}, Hsi-Wei and {Yoshida}, Tomohiro C.},
        title = "{exoALMA. V. Gaseous Emission Surfaces and Temperature Structures}",
      journal = {\apjl},
         year = 2025,
        month = may,
       volume = {984},
       number = {1},
          eid = {L10},
        pages = {L10},
          doi = {10.3847/2041-8213/adc437},
archivePrefix = {arXiv},
       eprint = {2504.19902},
 primaryClass = {astro-ph.EP},
       adsurl = {https://ui.adsabs.harvard.edu/abs/2025ApJ...984L..10G}
}

@ARTICLE{Teague_ea_2022,
       author = {{Teague}, Richard and {Bae}, Jaehan and {Benisty}, Myriam and {Andrews}, Sean M. and {Facchini}, Stefano and {Huang}, Jane and {Wilner}, David},
        title = "{Gas and Dust Shadows in the TW Hydrae Disk}",
      journal = {\apj},
         year = 2022,
        month = may,
       volume = {930},
       number = {2},
          eid = {144},
        pages = {144},
          doi = {10.3847/1538-4357/ac67a3},
archivePrefix = {arXiv},
       eprint = {2204.06691},
 primaryClass = {astro-ph.EP},
       adsurl = {https://ui.adsabs.harvard.edu/abs/2022ApJ...930..144T}
}

@ARTICLE{Benisty_ea_2018,
       author = {{Benisty}, M. and {Juh{\'a}sz}, A. and {Facchini}, S. and {Pinilla}, P. and {de Boer}, J. and {P{\'e}rez}, L.~M. and {Keppler}, M. and {Muro-Arena}, G. and {Villenave}, M. and {Andrews}, S. and {Dominik}, C. and {Dullemond}, C.~P. and {Gallenne}, A. and {Garufi}, A. and {Ginski}, C. and {Isella}, A.},
        title = "{Shadows and asymmetries in the T Tauri disk HD 143006: evidence for a misaligned inner disk}",
      journal = {\aap},
         year = 2018,
        month = nov,
       volume = {619},
          eid = {A171},
        pages = {A171},
          doi = {10.1051/0004-6361/201833913},
archivePrefix = {arXiv},
       eprint = {1809.01082},
 primaryClass = {astro-ph.EP},
       adsurl = {https://ui.adsabs.harvard.edu/abs/2018A&A...619A.171B}
}

@ARTICLE{Stadler_ea_2023,
       author = {{Stadler}, J. and {Benisty}, M. and {Izquierdo}, A. and {Facchini}, S. and {Teague}, R. and {Kurtovic}, N. and {Pinilla}, P. and {Bae}, J. and {Ansdell}, M. and {Loomis}, R. and {Mayama}, S. and {Perez}, L.~M. and {Testi}, L.},
        title = "{A kinematically detected planet candidate in a transition disk}",
      journal = {\aap},
         year = 2023,
        month = feb,
       volume = {670},
          eid = {L1},
        pages = {L1},
          doi = {10.1051/0004-6361/202245381},
archivePrefix = {arXiv},
       eprint = {2301.01684},
 primaryClass = {astro-ph.EP},
       adsurl = {https://ui.adsabs.harvard.edu/abs/2023A&A...670L...1S}
}

@ARTICLE{Hilder_ea_2025,
       author = {{Hilder}, Thomas and {Casey}, Andrew R. and {Price}, Daniel J. and {Pinte}, Christophe and {Izquierdo}, Andr{\'e}s F. and {Hardiman}, Caitlyn and {Bae}, Jaehan and {Barraza-Alfaro}, Marcelo and {Benisty}, Myriam and {Cataldi}, Gianni and {Curone}, Pietro and {Czekala}, Ian and {Facchini}, Stefano and {Fasano}, Daniele and {Flock}, Mario and {Fukagawa}, Misato and {Galloway-Sprietsma}, Maria and {Garg}, Himanshi and {Hall}, Cassandra and {Hammond}, Iain and {Huang}, Jane and {Ilee}, John D. and {Kanagawa}, Kazuhiro and {Lesur}, Geoffroy and {Longarini}, Cristiano and {Loomis}, Ryan and {Orihara}, Ryuta and {Rosotti}, Giovanni and {Stadler}, Jochen and {Teague}, Richard and {Yen}, Hsi-Wei and {Wafflard}, Gaylor and {Winter}, Andrew J. and {W{\"o}lfer}, Lisa and {Yoshida}, Tomohiro C. and {Zawadzki}, Brianna},
        title = "{exoALMA. VIII. Probabilistic Moment Maps and Data Products Using Nonparametric Linear Models}",
      journal = {\apjl},
         year = 2025,
        month = may,
       volume = {984},
       number = {1},
          eid = {L13},
        pages = {L13},
          doi = {10.3847/2041-8213/adc435},
archivePrefix = {arXiv},
       eprint = {2504.19416},
 primaryClass = {astro-ph.EP},
       adsurl = {https://ui.adsabs.harvard.edu/abs/2025ApJ...984L..13H}
}

@ARTICLE{Winter_ea_2025,
       author = {{Winter}, Andrew J. and {Benisty}, Myriam and {Izquierdo}, Andr{\'e}s F. and {Lodato}, Giuseppe and {Teague}, Richard and {Kimmig}, Carolin N. and {Andrews}, Sean M. and {Bae}, Jaehan and {Barraza-Alfaro}, Marcelo and {Cuello}, Nicol{\'a}s and {Curone}, Pietro and {Czekala}, Ian and {Facchini}, Stefano and {Fasano}, Daniele and {Hall}, Cassandra and {Hardiman}, Caitlyn and {Hilder}, Thomas and {Ilee}, John D. and {Fukagawa}, Misato and {Longarini}, Cristiano and {M{\'e}nard}, Fran{\c{c}}ois and {Orihara}, Ryuta and {Pinte}, Christophe and {Price}, Daniel J. and {Rosotti}, Giovanni and {Stadler}, Jochen and {Wilner}, David J. and {W{\"o}lfer}, Lisa and {Yen}, Hsi-Wei and {Yoshida}, Tomohiro C. and {Zawadzki}, Brianna},
        title = "{exoALMA. XVIII. Interpreting Large-scale Kinematic Structures as Moderate Warping}",
      journal = {\apjl},
         year = 2025,
        month = sep,
       volume = {990},
       number = {1},
          eid = {L10},
        pages = {L10},
          doi = {10.3847/2041-8213/adf113},
archivePrefix = {arXiv},
       eprint = {2507.11669},
 primaryClass = {astro-ph.EP},
       adsurl = {https://ui.adsabs.harvard.edu/abs/2025ApJ...990L..10W}
}

@ARTICLE{GoFish,
       author = {{Teague}, Richard},
        title = "{GoFish: Fishing for Line Observations in Protoplanetary Disks}",
      journal = {The Journal of Open Source Software},
         year = 2019,
        month = sep,
       volume = {4},
       number = {41},
          eid = {1632},
        pages = {1632},
          doi = {10.21105/joss.01632},
       adsurl = {https://ui.adsabs.harvard.edu/abs/2019JOSS....4.1632T}
}

@ARTICLE{Izquierdo_ea_2025,
       author = {{Izquierdo}, Andr{\'e}s F. and {Stadler}, Jochen and {Galloway-Sprietsma}, Maria and {Benisty}, Myriam and {Pinte}, Christophe and {Bae}, Jaehan and {Teague}, Richard and {Facchini}, Stefano and {W{\"o}lfer}, Lisa and {Longarini}, Cristiano and {Curone}, Pietro and {Andrews}, Sean M. and {Barraza-Alfaro}, Marcelo and {Cataldi}, Gianni and {Cuello}, Nicol{\'a}s and {Czekala}, Ian and {Fasano}, Daniele and {Flock}, Mario and {Fukagawa}, Misato and {Garg}, Himanshi and {Hall}, Cassandra and {Hammond}, Iain and {Hilder}, Thomas and {Huang}, Jane and {Ilee}, John D. and {Isella}, Andrea and {Kanagawa}, Kazuhiro and {Lesur}, Geoffroy and {Lodato}, Giuseppe and {Loomis}, Ryan A. and {Orihara}, Ryuta and {Price}, Daniel J. and {Rosotti}, Giovanni and {Testi}, Leonardo and {Yen}, Hsi-Wei and {Wafflard-Fernandez}, Gaylor and {Wilner}, David J. and {Winter}, Andrew J. and {Yoshida}, Tomohiro C. and {Zawadzki}, Brianna},
        title = "{exoALMA. III. Line-intensity Modeling and System Property Extraction from Protoplanetary Disks}",
      journal = {\apjl},
         year = 2025,
        month = may,
       volume = {984},
       number = {1},
          eid = {L8},
        pages = {L8},
          doi = {10.3847/2041-8213/adc439},
archivePrefix = {arXiv},
       eprint = {2504.19986},
 primaryClass = {astro-ph.EP},
       adsurl = {https://ui.adsabs.harvard.edu/abs/2025ApJ...984L...8I}
}

@ARTICLE{Izquierdo_ea_2026,
       author = {{Izquierdo}, Andr{\'e}s F. and {Bae}, Jaehan and {Facchini}, Stefano and {van Dishoeck}, Ewine F. and {Barraza-Alfaro}, Marcelo and {Benisty}, Myriam and {Teague}, Richard and {Stadler}, Jochen and {Andrews}, Sean M. and {Cataldi}, Gianni and {Cuello}, Nicol{\'a}s and {Curone}, Pietro and {Czekala}, Ian and {Fasano}, Daniele and {Flock}, Mario and {Fukagawa}, Misato and {Galloway-Sprietsma}, Maria and {Hall}, Cassandra and {Huang}, Jane and {Ilee}, John D. and {Isella}, Andrea and {Lawrence}, Jensen and {Lesur}, Geoffroy and {Lodato}, Giuseppe and {Longarini}, Cristiano and {Loomis}, Ryan A. and {M{\'e}nard}, Fran{\c{c}}ois and {Pinte}, Christophe and {Price}, Daniel J. and {Rosotti}, Giovanni and {Testi}, Leonardo and {Wilner}, David J. and {Winter}, Andrew J. and {W{\"o}lfer}, Lisa and {Zawadzki}, Brianna},
        title = "{exoALMA. XX. Tomographic Detection of Embedded Planets in Protoplanetary Disks}",
      journal = {\apjl},
         year = 2026,
        month = mar,
       volume = {1000},
       number = {1},
          eid = {L13},
        pages = {L13},
          doi = {10.3847/2041-8213/ae48fd},
archivePrefix = {arXiv},
       eprint = {2603.13157},
 primaryClass = {astro-ph.EP},
       adsurl = {https://ui.adsabs.harvard.edu/abs/2026ApJ..1000L..13I}
}

@ARTICLE{Casassus_Perez_2019,
       author = {{Casassus}, Simon and {P{\'e}rez}, Sebasti{\'a}n},
        title = "{Kinematic Detections of Protoplanets: A Doppler Flip in the Disk of HD 100546}",
      journal = {\apjl},
         year = 2019,
        month = oct,
       volume = {883},
       number = {2},
          eid = {L41},
        pages = {L41},
          doi = {10.3847/2041-8213/ab4425},
archivePrefix = {arXiv},
       eprint = {1906.06302},
 primaryClass = {astro-ph.EP},
       adsurl = {https://ui.adsabs.harvard.edu/abs/2019ApJ...883L..41C}
}

@ARTICLE{Teague_Foreman-Mackey_2018,
       author = {{Teague}, Richard and {Foreman-Mackey}, Daniel},
        title = "{A Robust Method to Measure Centroids of Spectral Lines}",
      journal = {Research Notes of the American Astronomical Society},
         year = 2018,
        month = sep,
       volume = {2},
       number = {3},
          eid = {173},
        pages = {173},
          doi = {10.3847/2515-5172/aae265},
archivePrefix = {arXiv},
       eprint = {1809.10295},
 primaryClass = {astro-ph.IM},
       adsurl = {https://ui.adsabs.harvard.edu/abs/2018RNAAS...2..173T}
}

@ARTICLE{Teague_ea_2026,
       author = {{Teague}, Richard and {Andrews}, Sean M. and {Bae}, Jaehan and {Barraza-Alfaro}, Marcelo and {Bodner}, Abigail and {Henshaw}, Jonathan D. and {Rosotti}, Giovanni},
        title = "{Using Structure Functions to Characterize Protoplanetary Disk Velocity Residuals}",
      journal = {\apj, submitted},
         year = 2026
}

@ARTICLE{Fukagawa_ea_2026,
       author = {{Fukagawa}, Misato and {Izquierdo}, Andr{\'e}s F. and {Stadler}, Jochen and {W{\"o}lfer}, Lisa and {Galloway-Sprietsma}, Maria and {Orihara}, Ryuta and {Aizawa}, Masataka and {Momose}, Munetake and {Fasano}, Daniele and {Benisty}, Myriam and {Teague}, Richard and {Facchini}, Stefano and {Pinte}, Christophe and {Andrews}, Sean M. and {Bae}, Jaehan and {Barraza-Alfaro}, Marcelo and {Cataldi}, Gianni and {Curone}, Pietro and {Czekala}, Ian and {Flock}, Mario and {Garg}, Himanshi and {Hall}, Cassandra and {Huang}, Jane and {Ilee}, John D. and {Lawrence}, Jensen and {Lesur}, Geoffroy and {Lodato}, Giuseppe and {Longarini}, Cristiano and {Loomis}, Ryan A. and {M{\'e}nard}, Fran{\c{c}}ois and {Price}, Daniel J. and {Rosotti}, Giovanni and {Yen}, Hsi-Wei and {Yoshida}, Tomohiro C. and {Wafflard-Fernandez}, Gaylor and {Wilner}, David J. and {Winter}, Andrew J. and {Zawadzki}, Brianna},
        title = "{exoALMA XXII: A Two-dimensional Atlas of Deviations from Keplerian Disks}",
      journal = {\apjl},
         year = 2026,
        month = mar,
       volume = {1000},
       number = {1},
          eid = {L15},
        pages = {L15},
          doi = {10.3847/2041-8213/ae3dc2},
archivePrefix = {arXiv},
       eprint = {2603.13015},
 primaryClass = {astro-ph.EP},
       adsurl = {https://ui.adsabs.harvard.edu/abs/2026ApJ..1000L..15F}
}

@ARTICLE{Speedie_ea_2024,
       author = {{Speedie}, Jessica and {Dong}, Ruobing and {Hall}, Cassandra and {Longarini}, Cristiano and {Veronesi}, Benedetta and {Paneque-Carre{\~n}o}, Teresa and {Lodato}, Giuseppe and {Tang}, Ya-Wen and {Teague}, Richard and {Hashimoto}, Jun},
        title = "{Gravitational instability in a planet-forming disk}",
      journal = {\nat},
         year = 2024,
        month = sep,
       volume = {633},
       number = {8028},
        pages = {58-62},
          doi = {10.1038/s41586-024-07877-0},
archivePrefix = {arXiv},
       eprint = {2409.02196},
 primaryClass = {astro-ph.EP},
       adsurl = {https://ui.adsabs.harvard.edu/abs/2024Natur.633...58S}
}

@ARTICLE{emcee,
       author = {{Foreman-Mackey}, Daniel and {Hogg}, David W. and {Lang}, Dustin and {Goodman}, Jonathan},
        title = "{emcee: The MCMC Hammer}",
      journal = {\pasp},
         year = 2013,
        month = mar,
       volume = {125},
       number = {925},
        pages = {306},
          doi = {10.1086/670067},
archivePrefix = {arXiv},
       eprint = {1202.3665},
 primaryClass = {astro-ph.IM},
       adsurl = {https://ui.adsabs.harvard.edu/abs/2013PASP..125..306F}
}

@ARTICLE{eddy,
       author = {{Teague}, Richard},
        title = "{eddy: Extracting Protoplanetary Disk Dynamics with Python}",
      journal = {The Journal of Open Source Software},
         year = 2019,
        month = feb,
       volume = {4},
       number = {34},
        pages = {1220},
          doi = {10.21105/joss.01220},
       adsurl = {https://ui.adsabs.harvard.edu/abs/2019JOSS....4.1220T}
}

@ARTICLE{Barraza-Alfaro_ea_2025,
       author = {{Barraza-Alfaro}, Marcelo and {Flock}, Mario and {B{\'e}thune}, William and {Teague}, Richard and {Bae}, Jaehan and {Benisty}, Myriam and {Cataldi}, Gianni and {Curone}, Pietro and {Czekala}, Ian and {Facchini}, Stefano and {Fasano}, Daniele and {Fukagawa}, Misato and {Galloway-Sprietsma}, Maria and {Garg}, Himanshi and {Hall}, Cassandra and {Huang}, Jane and {Ilee}, John D. and {Izquierdo}, Andr{\'e}s F. and {Kanagawa}, Kazuhiro and {Koch}, Eric W. and {Lesur}, Geoffroy and {Longarini}, Cristiano and {Loomis}, Ryan A. and {Orihara}, Ryuta and {Pinte}, Christophe and {Price}, Daniel J. and {Rosotti}, Giovanni and {Stadler}, Jochen and {Wafflard-Fernandez}, Gaylor and {Winter}, Andrew J. and {W{\"o}lfer}, Lisa and {Yen}, Hsi-Wei and {Yoshida}, Tomohiro C. and {Zawadzki}, Brianna},
        title = "{exoALMA. XVI. Predicting Signatures of Large-scale Turbulence in Protoplanetary Disks}",
      journal = {\apjl},
         year = 2025,
        month = may,
       volume = {984},
       number = {1},
          eid = {L21},
        pages = {L21},
          doi = {10.3847/2041-8213/adc42d},
archivePrefix = {arXiv},
       eprint = {2504.19853},
 primaryClass = {astro-ph.EP},
       adsurl = {https://ui.adsabs.harvard.edu/abs/2025ApJ...984L..21B}
}

@ARTICLE{Teague_ea_2025,
       author = {{Teague}, Richard and {Benisty}, Myriam and {Facchini}, Stefano and {Fukagawa}, Misato and {Pinte}, Christophe and {Andrews}, Sean M. and {Bae}, Jaehan and {Barraza-Alfaro}, Marcelo and {Cataldi}, Gianni and {Cuello}, Nicol{\'a}s and {Curone}, Pietro and {Czekala}, Ian and {Fasano}, Daniele and {Flock}, Mario and {Galloway-Sprietsma}, Maria and {Garg}, Himanshi and {Hall}, Cassandra and {Hammond}, Iain and {Hilder}, Thomas and {Huang}, Jane and {Ilee}, John D. and {Izquierdo}, Andr{\'e}s F. and {Kanagawa}, Kazuhiro and {Lesur}, Geoffroy and {Lodato}, Giuseppe and {Longarini}, Cristiano and {Loomis}, Ryan A. and {Masset}, Fr{\'e}d{\'e}ric and {Menard}, Francois and {Orihara}, Ryuta and {Price}, Daniel J. and {Rosotti}, Giovanni and {Stadler}, Jochen and {Testi}, Leonardo and {Yen}, Hsi-Wei and {Wafflard-Fernandez}, Gaylor and {Wilner}, David J. and {Winter}, Andrew J. and {W{\"o}lfer}, Lisa and {Yoshida}, Tomohiro C. and {Zawadzki}, Brianna},
        title = "{exoALMA. I. Science Goals, Project Design, and Data Products}",
      journal = {\apjl},
         year = 2025,
        month = may,
       volume = {984},
       number = {1},
          eid = {L6},
        pages = {L6},
          doi = {10.3847/2041-8213/adc43b},
archivePrefix = {arXiv},
       eprint = {2504.18688},
 primaryClass = {astro-ph.EP},
       adsurl = {https://ui.adsabs.harvard.edu/abs/2025ApJ...984L...6T}
}

@ARTICLE{Kolmogorov_1941,
       author = {{Kolmogorov}, Andrey Nikolaevich},
        title = "{The Local Structure of Turbulence in Incompressible Viscous Fluid for Very Large Reynolds' Numbers}",
      journal = {Doklady Akademii Nauk SSSR},
         year = 1941,
       volume = {30},
        pages = {301-305},
       adsurl = {https://ui.adsabs.harvard.edu/abs/1941DoSSR..30..301K}
}

@BOOK{Monin_Yaglom_1975,
       author = {{Monin}, A.~S. and {Yaglom}, A.~M.},
        title = "{Statistical Fluid Mechanics: Mechanics of Turbulence, Volume 2}",
    publisher = {MIT Press},
      address = {Cambridge, MA},
         year = 1975
}

@ARTICLE{Larson_1981,
       author = {{Larson}, R.~B.},
        title = "{Turbulence and star formation in molecular clouds}",
      journal = {\mnras},
         year = 1981,
        month = apr,
       volume = {194},
        pages = {809-826},
          doi = {10.1093/mnras/194.4.809},
       adsurl = {https://ui.adsabs.harvard.edu/abs/1981MNRAS.194..809L}
}

@ARTICLE{Ossenkopf_MacLow_2002,
       author = {{Ossenkopf}, V. and {Mac Low}, M.-M.},
        title = "{Turbulent velocity structure in molecular clouds}",
      journal = {\aap},
         year = 2002,
        month = jul,
       volume = {390},
        pages = {307-326},
          doi = {10.1051/0004-6361:20020629},
       adsurl = {https://ui.adsabs.harvard.edu/abs/2002A&A...390..307O}
}

@ARTICLE{Heyer_Brunt_2004,
       author = {{Heyer}, Mark H. and {Brunt}, Christopher M.},
        title = "{The Universality of Turbulence in Galactic Molecular Clouds}",
      journal = {\apjl},
         year = 2004,
        month = nov,
       volume = {615},
       number = {1},
        pages = {L45-L48},
          doi = {10.1086/425978},
       adsurl = {https://ui.adsabs.harvard.edu/abs/2004ApJ...615L..45H}
}

@ARTICLE{Elmegreen_Scalo_2004,
       author = {{Elmegreen}, Bruce G. and {Scalo}, John},
        title = "{Interstellar Turbulence I: Observations and Processes}",
      journal = {\araa},
         year = 2004,
        month = sep,
       volume = {42},
       number = {1},
        pages = {211-273},
          doi = {10.1146/annurev.astro.41.011802.094859},
       adsurl = {https://ui.adsabs.harvard.edu/abs/2004ARA&A..42..211E}
}

@ARTICLE{Esquivel_Lazarian_2005,
       author = {{Esquivel}, A. and {Lazarian}, A.},
        title = "{Velocity Centroids as Tracers of the Turbulent Velocity Statistics}",
      journal = {\apj},
         year = 2005,
        month = sep,
       volume = {631},
       number = {1},
        pages = {320-350},
          doi = {10.1086/432458},
archivePrefix = {arXiv},
       eprint = {astro-ph/0401603},
 primaryClass = {astro-ph},
       adsurl = {https://ui.adsabs.harvard.edu/abs/2005ApJ...631..320E}
}

@ARTICLE{Bruno_Carbone_2013,
       author = {{Bruno}, Roberto and {Carbone}, Vincenzo},
        title = "{The Solar Wind as a Turbulence Laboratory}",
      journal = {Living Reviews in Solar Physics},
         year = 2013,
        month = dec,
       volume = {10},
       number = {1},
          eid = {2},
        pages = {2},
          doi = {10.12942/lrsp-2013-2},
       adsurl = {https://ui.adsabs.harvard.edu/abs/2013LRSP...10....2B}
}

@ARTICLE{Jorsater_vanMoorsel_1995,
       author = {{J{\"o}rs{\"a}ter}, S. and {van Moorsel}, G. A.},
        title = "{High Resolution Neutral Hydrogen Observations of the Barred Spiral Galaxy NGC 1365}",
      journal = {\aj},
         year = 1995,
       volume = {110},
        pages = {2037},
       adsurl = {https://ui.adsabs.harvard.edu/abs/1995AJ....110.2037J}
}

@ARTICLE{Czekala_ea_2021,
       author = {{Czekala}, Ian and {Loomis}, Ryan A. and {Teague}, Richard and {Booth}, Alice S. and {Huang}, Jane and {Cataldi}, Gianni and {Ilee}, John D. and {Law}, Charles J. and {Walsh}, Catherine and {Bosman}, Arthur D. and {Guzm{\'a}n}, Viviana V. and {Le Gal}, Romane and {{\"O}berg}, Karin I. and {Yamato}, Yoshihide and {Aikawa}, Yuri and {Andrews}, Sean M. and {Bae}, Jaehan and {Bergin}, Edwin A. and {Bergner}, Jennifer B. and {Cleeves}, L. Ilsedore and {Kurtovic}, Nicolas T. and {M{\'e}nard}, Fran{\c{c}}ois and {Nomura}, Hideko and {P{\'e}rez}, Laura M. and {Qi}, Chunhua and {Schwarz}, Kamber R. and {Tsukagoshi}, Takashi and {Waggoner}, Abygail R. and {Wilner}, David J. and {Zhang}, Ke},
        title = "{Molecules with ALMA at Planet-forming Scales (MAPS). II. CLEAN Strategies for Synthesizing Images of Molecular Line Emission in Protoplanetary Disks}",
      journal = {\apjs},
         year = 2021,
        month = nov,
       volume = {257},
       number = {1},
          eid = {2},
        pages = {2},
          doi = {10.3847/1538-4365/ac1430},
archivePrefix = {arXiv},
       eprint = {2109.06188},
 primaryClass = {astro-ph.EP},
       adsurl = {https://ui.adsabs.harvard.edu/abs/2021ApJS..257....2C}
}

@ARTICLE{Wolfer_ea_2025,
       author = {{W{\"o}lfer}, Lisa and {Barraza-Alfaro}, Marcelo and {Teague}, Richard and {Curone}, Pietro and {Benisty}, Myriam and {Fukagawa}, Misato and {Bae}, Jaehan and {Cataldi}, Gianni and {Czekala}, Ian and {Facchini}, Stefano and {Fasano}, Daniele and {Flock}, Mario and {Galloway-Sprietsma}, Maria and {Garg}, Himanshi and {Hall}, Cassandra and {Huang}, Jane and {Ilee}, John D. and {Izquierdo}, Andr{\'e}s F. and {Kanagawa}, Kazuhiro and {Lesur}, Geoffroy and {Longarini}, Cristiano and {Loomis}, Ryan A. and {M{\'e}nard}, Fran{\c{c}}ois and {Nath}, Anika and {Orihara}, Ryuta and {Pinte}, Christophe and {Price}, Daniel J. and {Rosotti}, Giovanni and {Stadler}, Jochen and {Wafflard-Fernandez}, Gaylor and {Winter}, Andrew J. and {Yen}, Hsi-Wei and {Yoshida}, Tomohiro C. and {Zawadzki}, Brianna},
        title = "{exoALMA. XVII. Characterizing the Gas Dynamics around Dust Asymmetries}",
      journal = {\apjl},
         year = 2025,
        month = may,
       volume = {984},
       number = {1},
          eid = {L22},
        pages = {L22},
          doi = {10.3847/2041-8213/adc42c},
archivePrefix = {arXiv},
       eprint = {2504.20023},
 primaryClass = {astro-ph.EP},
       adsurl = {https://ui.adsabs.harvard.edu/abs/2025ApJ...984L..22W}
}

@ARTICLE{Benisty_ea_2026,
       author = {{Benisty}, Myriam and {Izquierdo}, Andr{\'e}s F. and {Stadler}, Jochen and {Galloway-Sprietsma}, Maria and {Facchini}, Stefano and {Winter}, Andrew J. and {Bae}, Jaehan and {Fukagawa}, Misato and {Teague}, Richard and {Pinte}, Christophe and {Andrews}, Sean M. and {Barraza-Alfaro}, Marcelo and {Cataldi}, Gianni and {Curone}, Pietro and {Czekala}, Ian and {Fasano}, Daniele and {Flock}, Mario and {Garg}, Himanshi and {Huang}, Jane and {Ilee}, John D. and {Kanagawa}, Kazuhiro and {Lawrence}, Jensen and {Lesur}, Geoffroy and {Lodato}, Giuseppe and {Longarini}, Cristiano and {Loomis}, Ryan A. and {M{\'e}nard}, Fran{\c{c}}ois and {Orihara}, Ryuta and {Price}, Daniel J. and {Rosotti}, Giovanni and {Wafflard-Fernandez}, Gaylor and {Wilner}, David J. and {W{\"o}lfer}, Lisa and {Yen}, Hsi-Wei and {Yoshida}, Tomohiro C. and {Zawadzki}, Brianna},
        title = "{exoALMA. XXI. The Morphology and Dynamics of Vertical Flows}",
      journal = {\apjl},
         year = 2026,
        month = mar,
          doi = {10.3847/2041-8213/ae360e},
archivePrefix = {arXiv},
       eprint = {2603.12861},
 primaryClass = {astro-ph.EP}
}

@ARTICLE{Flock_ea_2011,
       author = {{Flock}, Mario and {Dzyurkevich}, Natalia and {Klahr}, Hubert and {Turner}, Neal J. and {Henning}, Thomas},
        title = "{Turbulence and Steady Flows in Three-dimensional Global Stratified Magnetohydrodynamic Simulations of Accretion Disks}",
      journal = {\apj},
         year = 2011,
        month = jul,
       volume = {735},
       number = {2},
          eid = {122},
        pages = {122},
          doi = {10.1088/0004-637X/735/2/122},
archivePrefix = {arXiv},
       eprint = {1104.4565},
 primaryClass = {astro-ph.SR},
       adsurl = {https://ui.adsabs.harvard.edu/abs/2011ApJ...735..122F}
}

@ARTICLE{astropy2013,
       author = {{Astropy Collaboration} and others},
        title = "{Astropy: A community Python package for astronomy}",
      journal = {\aap},
         year = 2013,
        month = oct,
       volume = {558},
        pages = {A33},
          doi = {10.1051/0004-6361/201322068},
       adsurl = {https://ui.adsabs.harvard.edu/abs/2013A&A...558A..33A}
}

@ARTICLE{astropy2018,
       author = {{Astropy Collaboration} and others},
        title = "{The Astropy Project: Building an Open-science Project and Status of the v2.0 Core Package}",
      journal = {\aj},
         year = 2018,
        month = sep,
       volume = {156},
       number = {3},
        pages = {123},
          doi = {10.3847/1538-3881/aabc4f},
       adsurl = {https://ui.adsabs.harvard.edu/abs/2018AJ....156..123A}
}

@ARTICLE{astropy2022,
       author = {{Astropy Collaboration} and others},
        title = "{The Astropy Project: Sustaining and Growing a Community-oriented Open-source Project and the Latest Major Release (v5.0) of the Core Package}",
      journal = {\apj},
         year = 2022,
        month = aug,
       volume = {935},
       number = {2},
        pages = {167},
          doi = {10.3847/1538-4357/ac7c74},
       adsurl = {https://ui.adsabs.harvard.edu/abs/2022ApJ...935..167A}
}

@ARTICLE{matplotlib,
       author = {{Hunter}, John D.},
        title = "{Matplotlib: A 2D Graphics Environment}",
      journal = {Computing in Science and Engineering},
         year = 2007,
        month = may,
       volume = {9},
       number = {3},
        pages = {90-95},
          doi = {10.1109/MCSE.2007.55},
       adsurl = {https://ui.adsabs.harvard.edu/abs/2007CSE.....9...90H}
}

@ARTICLE{numpy,
       author = {{Harris}, Charles R. and {Millman}, K. Jarrod and {van der Walt}, St{\'e}fan J. and others},
        title = "{Array programming with NumPy}",
      journal = {Nature},
         year = 2020,
        month = sep,
       volume = {585},
       number = {7825},
        pages = {357-362},
          doi = {10.1038/s41586-020-2649-2},
       adsurl = {https://ui.adsabs.harvard.edu/abs/2020Natur.585..357H}
}

@ARTICLE{scipy,
       author = {{Virtanen}, Pauli and {Gommers}, Ralf and {Oliphant}, Travis E. and others},
        title = "{SciPy 1.0: fundamental algorithms for scientific computing in Python}",
      journal = {Nature Methods},
         year = 2020,
        month = feb,
       volume = {17},
        pages = {261-272},
          doi = {10.1038/s41592-019-0686-2},
       adsurl = {https://ui.adsabs.harvard.edu/abs/2020NatMe..17..261V}
}

@ARTICLE{Lynden-Bell_Kalnajs_1972,
       author = {{Lynden-Bell}, D. and {Kalnajs}, A.~J.},
        title = "{On the generating mechanism of spiral structure}",
      journal = {\mnras},
         year = 1972,
       volume = {157},
        pages = {1},
          doi = {10.1093/mnras/157.1.1}
}

@ARTICLE{Pinte_ea_2018,
       author = {{Pinte}, C. and {Price}, D.~J. and {M{\'e}nard}, F. and {Duch{\^e}ne}, G. and {Dent}, W.~R.~F. and {Hill}, T. and {de Gregorio-Monsalvo}, I. and {Hales}, A. and {Mentiplay}, D.},
        title = "{Kinematic Evidence for an Embedded Protoplanet in a Circumstellar Disk}",
      journal = {\apjl},
         year = 2018,
        month = jun,
       volume = {860},
       number = {1},
          eid = {L13},
        pages = {L13},
          doi = {10.3847/2041-8213/aac6dc}
}

@ARTICLE{Pinte_ea_2019,
       author = {{Pinte}, C. and {van der Plas}, G. and {M{\'e}nard}, F. and {Price}, D.~J. and {Christiaens}, V. and {Hill}, T. and {Mentiplay}, D. and {Ginski}, C. and {Choquet}, E. and {Boehler}, Y. and {Duch{\^e}ne}, G. and {Perez}, S. and {Casassus}, S.},
        title = "{Kinematic detection of a planet carving a gap in a protoplanetary disk}",
      journal = {Nature Astronomy},
         year = 2019,
        month = aug,
       volume = {3},
        pages = {1109-1114},
          doi = {10.1038/s41550-019-0852-6}
}

@ARTICLE{Rosenfeld_ea_2014,
       author = {{Rosenfeld}, Katherine A. and {Chiang}, Eugene and {Andrews}, Sean M.},
        title = "{Fast Radial Flows in Transition Disk Holes}",
      journal = {\apj},
         year = 2014,
        month = jan,
       volume = {782},
       number = {2},
          eid = {62},
        pages = {62},
          doi = {10.1088/0004-637X/782/2/62}
}

@ARTICLE{Hall_ea_2020,
       author = {{Hall}, C. and {Dong}, R. and {Teague}, R. and {Terry}, J. and {Pinte}, C. and {Paneque-Carre{\~n}o}, T. and {Veronesi}, B. and {Alexander}, R.~D. and {Lodato}, G.},
        title = "{Predicting the Kinematic Evidence of Gravitational Instability}",
      journal = {\apj},
         year = 2020,
        month = nov,
       volume = {904},
       number = {2},
          eid = {148},
        pages = {148},
          doi = {10.3847/1538-4357/abac17}
}

@ARTICLE{Barraza-Alfaro_ea_2021,
       author = {{Barraza-Alfaro}, Marcelo and {Flock}, Mario and {Marino}, Sebastian and {P{\'e}rez}, Sebasti{\'a}n},
        title = "{Observability of the vertical shear instability in protoplanetary disk CO kinematics}",
      journal = {\aap},
         year = 2021,
       volume = {653},
          eid = {A113},
        pages = {A113},
          doi = {10.1051/0004-6361/202140535},
archivePrefix = {arXiv},
       eprint = {2106.01159}
}

@ARTICLE{Nelson_ea_2013,
       author = {{Nelson}, Richard P. and {Gressel}, Oliver and {Umurhan}, Orkan M.},
        title = "{Linear and non-linear evolution of the vertical shear instability in accretion discs}",
      journal = {\mnras},
         year = 2013,
       volume = {435},
       number = {3},
        pages = {2610-2632},
          doi = {10.1093/mnras/stt1475},
archivePrefix = {arXiv},
       eprint = {1209.2753}
}

@ARTICLE{Guan_ea_2009,
       author = {{Guan}, Xiaoyue and {Gammie}, Charles F. and {Simon}, Jacob B. and {Johnson}, Bryan M.},
        title = "{Locality of MHD Turbulence in Isothermal Disks}",
      journal = {\apj},
         year = 2009,
       volume = {694},
       number = {2},
        pages = {1010-1018},
          doi = {10.1088/0004-637X/694/2/1010},
archivePrefix = {arXiv},
       eprint = {0901.0273}
}

@ARTICLE{Goldreich_Lynden-Bell_1965,
       author = {{Goldreich}, P. and {Lynden-Bell}, D.},
        title = "{II. Spiral arms as sheared gravitational instabilities}",
      journal = {\mnras},
         year = 1965,
       volume = {130},
       number = {2},
        pages = {125-158},
          doi = {10.1093/mnras/130.2.125}
}

@ARTICLE{Heinemann_Papaloizou_2009a,
       author = {{Heinemann}, T. and {Papaloizou}, J.~C.~B.},
        title = "{The excitation of spiral density waves through turbulent fluctuations in accretion discs - I. WKBJ theory}",
      journal = {\mnras},
         year = 2009,
       volume = {397},
       number = {1},
        pages = {52-63},
          doi = {10.1111/j.1365-2966.2009.14799.x},
archivePrefix = {arXiv},
       eprint = {0812.2068}
}

@ARTICLE{Heinemann_Papaloizou_2009b,
       author = {{Heinemann}, T. and {Papaloizou}, J.~C.~B.},
        title = "{The excitation of spiral density waves through turbulent fluctuations in accretion discs - II. Numerical simulations with MRI-driven turbulence}",
      journal = {\mnras},
         year = 2009,
       volume = {397},
       number = {1},
        pages = {64-74},
          doi = {10.1111/j.1365-2966.2009.14800.x},
archivePrefix = {arXiv},
       eprint = {0812.2471}
}
\bibliographystyle{aasjournalv7}

\end{document}